\documentclass[english,pre]{revtex4}
\usepackage[T1]{fontenc}
\usepackage[latin1]{inputenc}
\usepackage{verbatim}
\usepackage{amsmath}
\usepackage{color}
\usepackage{graphicx}
\usepackage{amssymb}
\makeatletter

\begin{document}

\title{Chaotic Gene Regulatory Networks Can Be Robust Against Mutations and Noise}
\author{Volkan Sevim$^{1}$}
\altaffiliation[Current address: ]{Physics Dept., Box 90305, Duke University, Durham, NC 27708}
\email{volkan.sevim@duke.edu}
\author{Per Arne Rikvold$^{1,2}$}
\email{rikvold@scs.fsu.edu}

\affiliation{$^{1}$ School of Computational Science, Center for Materials Research
and Technology, and Department of Physics, Florida State University,
Tallahassee, FL 32306-4120, USA\\
 $^{2}$ National High Magnetic Field Laboratory, Tallahassee, FL
32310-3706, USA}

\date{\today}

\begin{abstract}
Robustness to mutations and noise has been shown to evolve through
stabilizing selection for optimal phenotypes in model gene regulatory
networks. The ability to evolve robust mutants is known to depend
on the network architecture. How do the dynamical properties and state-space
structures of networks with high and low robustness differ? Does selection
operate on the global dynamical behavior of the networks? What kind
of state-space structures are favored by selection? We provide damage
propagation analysis and an extensive statistical analysis of state
spaces of these model networks to show that the change in their dynamical
properties due to stabilizing selection for optimal phenotypes is minor.
Most notably, the networks that are most robust to both mutations
and noise are highly chaotic. Certain properties of chaotic networks,
such as being able to produce large attractor basins, can be useful
for maintaining a stable gene-expression pattern. Our findings indicate
that conventional measures of stability, such as the damage-propagation
rate, do not provide much information about robustness to mutations
or noise in model gene regulatory networks.
\end{abstract}

\maketitle

\section{Introduction}

\newcommand{\mx}{\mathbf}
\newcommand{\wmx}{\mathbf{W}}
\newcommand{\state}{{\mathbf{s}}}
\newcommand{\initst}{{\state}(0)}
\newcommand{\finst}{{{\state}{^*}}}
\newcommand{\sgn}{{\mathrm{sgn}}}
\newcommand{\stilde} {{\tilde s}}
\newcommand{\stildest} {\mathbf{\tilde s}}
\newcommand{\seta} {{\cal A}_t}
\newcommand{\setb} {{\cal B}_t}
\newcommand{\meank} {{\langle k\rangle}}
\newcommand{\Rmu} {R_{{\mathrm \mu}}}

The genetic architecture of biological organisms shows remarkable
robustness against both structural and environmental perturbations
\citep{RobustnessReview:2003,WagnerBook,Hansen:2006,Barkai:1997}.
For example, quantitative models indicate that the functionality of
the \emph{Drosophila} segment polarity gene network is extremely insensitive
to variations in initial conditions \citep{vondassow:2000} and robust
against architectural modifications \citep{Vondassow:2002}. (Also
see \citet{Chaves:2005,AlbertOthmer:2003}.) Gene knock-out studies
on yeast have shown that almost 40\% of the genes on chromosome V
have either negligible or no effects on the growth rate \citep{Smith:1996}.
Certain cellular networks, such as the \emph{E. Coli} chemotaxis network,
are also known to be very robust to variations in biochemical parameters
\citep{Barkai:1997,Alon:1999}. %
{}

The genetic regulatory networks that control the developmental dynamics
buffer perturbations and maintain a stable phenotype. That is why
phenotypic variation within most species is quite small, despite the
organisms being exposed to a wide range of environmental and genetic
perturbations \citep{WagnerBook}. It has been proposed that genetic
robustness evolved through stabilizing selection for a phenotypic
optimum \citep{Wagner:1996,Wagner:2000,RobustnessReview:2003}. \citet{Wagner:1996}
showed that this in fact can be true by modeling a developmental process
within an evolutionary scenario, in which the genetic interaction
sequence represents the development, and the stationary configuration
of the gene network represents the phenotype.
His results indicate that the genetic robustness of a population of
model genetic regulatory networks can gradually increase through stabilizing
selection, in which deviations from the {}``optimal'' stationary
state (phenotype) are considered deleterious \citep{Ciliberti:2007}. 

In this paper, we focus on the effects of evolution of genetic robustness
on the dynamics of gene regulatory networks in general. First, we
examine the relationship between genetic robustness and the dynamical
character of the networks. By dynamical character, we mean the stability
of the expression states of the network against small perturbations,
or noise. 

Models that are employed to study dynamics of gene networks, such
as random Boolean networks (RBN) \citep{Kauffman:1969,KAUF93,Aldana:2003,AldanaReview,DrosselRBNReview}
or some variants of random threshold networks (RTN) \citep{Kurten:1988a}
have been known to undergo a phase transition at low connectivities
\citep{Rohlf:2002,Luque:2001}, giving rise to change in dynamical
behavior: on average, small perturbations will percolate through the
network above the threshold connectivity (chaotic phase), whereas
they stay confined to a part of the network below the threshold (ordered
phase). Intuitively, one might expect to find that robustness to mutations  
(which are permanent structural changes, not dynamic perturbations)
is related to the dynamical behavior of the system, therefore, ordered gene 
regulatory networks should be genetically more robust than the chaotic ones. 
However, this is not necessarily true. Here, we show that 
the relation between the dynamical character
of a genetic regulatory network and its mutational robustness can
be quite the opposite. In fact, even earlier studies provide a clue on this issue:
for gene networks that have undergone selection
(i.e., evolved), mutational robustness is known to increase with increasing
connectivity \citep{Wagner:1996}. On the other hand, chaoticity has
been shown to increase with increasing connectivity in random gene
regulatory networks \citep{Rohlf:2002,AldanaReview}.
These facts seem to contradict the intuitive interpretation of robustness
since they suggest that chaotic networks can be mutationally robust. 

Although this inference is proven true in this paper, such an assessment cannot be based
on the previous studies of the dynamics of random networks \citep{AldanaReview,Rohlf:2002,KAUF93,Kurten:1988,Kurten:1988a,Luque:2001},
as the evolved networks (with high mutational robustness and connectivity)
mentioned above \citep{Wagner:1996} have undergone selection. The selection
process could potentially tune a network to exhibit a different dynamical
character than its random ancestors. Therefore, the dynamical character 
of the evolved networks should be studied independently and then compared
with their random counterparts.
Here, we study the dynamics of the evolved networks numerically and show that selection
for an optimal phenotype indeed has only a minor effect on their global dynamical
behavior. This result indicates that the evolution of mutational robustness
cannot be understood in terms of simple dynamical measures. 

We also provide statistics on robustness to \emph{noise}, which is
the ability of a network to reach its {}``optimal'' steady state
after a perturbation to the gene-expression \emph{trajectory}. Computer
simulations indicate that mutational robustness is correlated to robustness
of the gene-expression trajectory to small perturbations (noise),
at least for short trajectories. This result is supported by recent
studies \citep{Kaneko:2007PLOSONE,Ciliberti:2007}. For perturbations
of arbitrary magnitude, the basin size of the steady-state attractor
provides a better measure. Our analysis shows that basin sizes of
densely connected networks (which are highly chaotic) have a broad
distribution, and therefore such networks can have very large attractor
basins \citep{AldanaReview,Ganguli:2007,Shmulevich:2007}. The yeast cell-cycle network
has been reported to have similar properties \citep{Li:2006}. Although
chaoticity is just a side-effect of high connectivity, and not directly
selected for during evolution, it appears that this intrinsic property
of chaotic networks can be useful in terms of robustness to noise.
When all of these measures are taken into account, chaotic dynamics
does not seem be an obstacle for the networks with high connectivity
since they are more robust to both mutations and noise than the ones
that are sparsely connected.

The organization of the rest of this paper is as follows. We describe
the model in Sec.~\ref{sec:Model} and explain its implementation
in simulations in Sec.~\ref{sec:Methods}. We give the results in
Sec.~\ref{sec:Results} and discuss their implications in Sec~\ref{sec:Discussion}.


\section{Model\label{sec:Model}}

We use the model introduced by \citet{Wagner:1996}, which has
also been used with some modifications by other researchers \citep{Azevedo:2006,Siegal:2002,Bergman:2003}.
Each individual is represented by a regulatory gene network consisting
of $N$ genes. The expression level of each gene, $s_{i},$ can be
either $+1$ or $-1$, meaning that the gene is expressed or not,
respectively. The expression states change in time according to regulatory
interactions between the genes. The time development of the expression states 
(i.e., the dynamical trajectory taken by the network) represents a 
developmental pathway. This deterministic, discrete-time dynamics of the 
development is given by a set of nonlinear difference equations, 
\begin{equation}
s_{i}(t+1)=\left\{ \begin{array}{cc}
\mathrm{\mathbf{\mathrm{sgn}}}\left(\sum_{j=1}^{N}w_{ij}s_{j}(t)\right), & \sum_{j=1}^{N}w_{ij}s_{j}(t)\neq0\\
s_{i}(t), & \sum_{j=1}^{N}w_{ij}s_{j}(t)=0\end{array}\right.\;,\label{eq:main}\end{equation}
 where sgn is the sign function and $w_{ij}$ is the strength of the
influence of gene $j$ on gene $i$. Nonzero elements $w_{ij}$ of the $N\times N$
matrix $\wmx$ are independent random numbers drawn from a gaussian
distribution with zero mean and unit variance. The diagonal elements
of \textbf{$\wmx$} are allowed to be nonzero, corresponding to self-regulation.
The mean number of nonzero elements in $\wmx$ is controlled by the
connectivity density, $c$, which is the probability that any given $w_{ij}$
is nonzero. Thus, the mean degree of the network is $\left\langle k\right\rangle =cN.$
To clarify, $t$ represents the \emph{developmental} time, through which
genetic interactions occur. It is different from the \emph{evolutionary}
time, $T$, which will be explained in the next section. 

The dynamics given by Eq.~\eqref{eq:main} can display a wide variety
of features. For a specified initial state $\initst$, the
network reaches an attractor (either a fixed point or a limit
cycle) after a transient period. Transient time, number of attractors,
attractor periods, etc. can differ depending
on the connectivity of the network, and from one realization of $\wmx$
to another. The fitness of an individual is defined by whether
it can reach a developmental equilibrium, i.e., a fixed point, which
is a fixed gene-expression pattern, $\mathbf{s^{*}}$, in
a {}``reasonable'' transient time. (It has been shown
that selection for developmental stability is sufficient for evolution
of mutational robustness \citep{Siegal:2002}, i.e., deviations from
$\mathbf{s^{*}}$ do not have to be deleterious, as long as the 
gene-expression configuration reaches a fixed point. 
However, we shall adopt the criterion used by \citet{Wagner:1996}
for compatibility.) Further details of the model are explained in
the next section.

\section{Methods\label{sec:Methods}}

\subsection{Before evolution: Generation of networks}

We studied populations of random networks with $N=10$. We use these relatively small networks to be able to enumarate all network states exhaustively for a large set of realizations. In the simulations, each
network was first assigned a random interaction matrix $\wmx$ and an initial state
$\initst$. $\wmx$ was generated as follows. For each $w_{ij}$,
a random number uniformly distributed on $[0,1)$ was generated, and
$w_{ij}$ was set to zero if the random number was greater than the
connectivity density, $c$. Otherwise, $w_{ij}$ was assigned a random
number drawn from a gaussian distribution with zero mean and unit
variance. Then, each {}``gene'' of the initial configuration, $s_{i}(0)$,
was assigned either $-1$ or $+1$ at random, each with probability
1/2.

After $\wmx$ and $\initst$ were created, the developmental dynamics were started,
and the network's stability was evaluated. If the system reached a
fixed point, $\finst,$ in $3N$ time steps, then it was considered
\emph{viable} and kept. Otherwise it was considered unstable, both
$\wmx$ and $\initst$ were discarded, and the process was repeated 
until a viable network was generated. (A viable
network can have several fixed points and/or limit cycles in addition
to $\finst$.) For each viable network, its fixed point, $\mathbf{s^{*}}$,
was regarded as the {}``optimal'' gene-expression state of the system.
This is the only modification we made to the model used by \citet{Wagner:1996}:
we accept any $\finst$ as long as it can be reached within $3N$
time steps from $\initst$, whereas \citet{Wagner:1996} generated networks 
with preassigned random $\initst$ and $\mathbf{s^{*}}$
\footnote{One also needs to consider the Hamming distance, $h$, between $\initst$
and $\finst$ to assess mutational robustness of these networks \citep{Ciliberti:2007}.
We ignore this fact since producing networks with a predetermined $h$
is computationally not very feasible. The probability distribution of
$h$ for the networks we generate follows approximately a gaussian
with a mean between 2.8 and 4.5, increasing with $\meank$. See supplementary Fig. S2. %
}.

\subsection{Evolution}

In order to generate a collection of more robust networks, a mutation-selection
process was simulated for each viable network as follows. First, a
clan of $\mathcal{N}=500$ identical copies of each network was generated.
For each member of the clan, a four-step process was performed for
$T=400$ generations:

\begin{enumerate}
\item Recombination: Each pair of the $N$ rows of consecutive matrices
in the clan were swapped with probability 1/2. Since the networks
were already shuffled in step 4 (see below), there was no need to
pick random pairs. 
\item Mutation: Each nonzero $w_{ij}$ was replaced with probability $1/(cN^{2})$
by a new random number drawn from the same standard gaussian distribution.
Thus, on average, one matrix element was changed per matrix per generation. 
\item Fitness evaluation: Each network was run starting from the original
initial condition $\initst$. If the network reached a fixed point,
$\mathbf{s}^{\dagger},$ within $t=3N$ developmental time steps, then 
its fitness was calculated using
\begin{equation}
\mathbf{\mathrm{f}(\mathbf{\mathbf{s}^{\dagger},\finst})=\exp(-\mathrm{H}^{2}(\mathbf{\mathbf{s}^{\dagger},\finst})/\sigma_{{\rm s}}))}\;,
\label{eq:fitness}
\end{equation}
 where $\mathrm{H}(\mathbf{\mathbf{s}^{\dagger},\finst})$ denotes
the normalized Hamming distance between $\mathbf{s}^{\dagger}$ and
$\mathbf{\finst}$, $\sigma_{{\rm s}}$ denotes the inverse of the
strength of selection, $\finst$ is the optimal gene-expression state,
which is the final gene-expression state of the original network that
{}``founded'' the clan. We used $\sigma_{{\rm s}}=0.1$. If the
network could not reach a fixed point, it was assigned the minimum
nonzero fitness value, $\exp(-1/\sigma_{{\rm s}}).$ 
\item Selection/Asexual Reproduction: The fitness of each network was normalized
to the fitness value of the most fit network in the clan. Then a
network was chosen at random and duplicated into the descendant clan
with probability equal to its normalized fitness. This process was
repeated until the size of the descendant clan reached $\mathcal{N}$.
Then the old clan was discarded, and the descendant clan was kept
as the next generation. This process allows multiple copies
(offspring) of the same network to appear in the descendant clan,
while some networks may not be propagated to the next generation due to
genetic drift. 
\end{enumerate}
At the end of the $T=400$ generation (evolutionary time) selection,
any unstable networks were removed from the evolved clan. 

Some steps of the process above may be unnecessary, and in fact,
the results do not depend strongly on model details \citep{Wagner:1996,HuertaSanchez:2007}.
Nevertheless, the entire procedure of \citet{Wagner:1996} was retained
for compatibility.

\subsection{Assessment of Robustness \label{sub:Assesment-of-Epigenetic}}

The mutational robustness, $R_{{\rm \mu}},$ of a network was assessed
as follows. First, a nonzero $w_{ij}$ was picked at random and replaced
by a new random number with the same standard gaussian distribution.
Then, the developmental dynamics were started, and it was checked whether the system
reached the same stationary state, $\finst$, within $t=3N$ time
steps. This process was repeated $5000c$ times, starting from the original
matrix. The robustness of the original network before evolution
was defined as the fraction of singly-mutated networks that reached
$\finst$. For the evolved networks, we picked one sample network
at random from the clan and used the same procedure to assess its
mutational robustness. 

\begin{figure}
\centering
\includegraphics[width=0.5\textwidth]{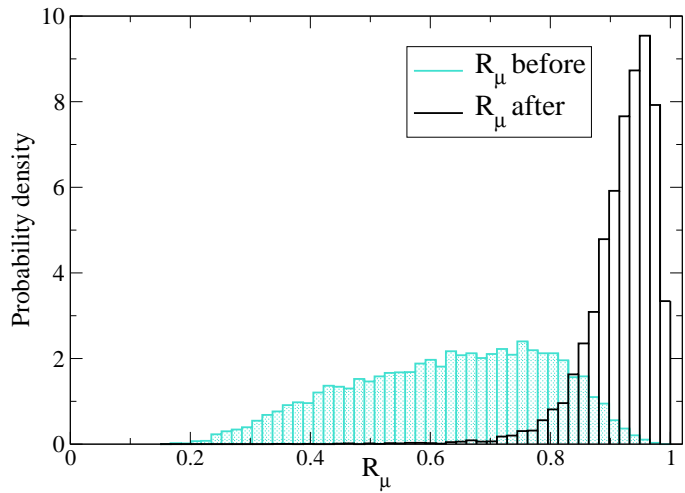}

\caption{Probability density of mutational robustness ($R_{{\rm \mu}}$)
of 10000 sample networks before (filled bars) and after (empty bars)
evolution with $N=10$ and $\left\langle k\right\rangle =Nc=10$.
Before: $\left\langle R_{{\rm \mu}}\right\rangle =0.63,$ $\sigma=0.16$.
After: $\left\langle R_{{\rm \mu}}\right\rangle =0.92,$ $\sigma=0.06$.
The evolved distribution was calculated by sampling one network from
each of 10000 evolved clans. The mean indegree, $\left\langle k\right\rangle$, 
rather than the connectivity density, $c$, is the parameter that controls
the behavior of the system \citep{AldanaReview,DrosselRBNReview}.
See Sec. \ref{sec:Methods} for other model parameters.}

\label{fig:symmetry-and-robustness}
\end{figure}

\section{Results\label{sec:Results} }

The stabilizing selection described above increases the robustness
of a model population of gene networks against mutations \citep{Wagner:1996}.
Figure \ref{fig:symmetry-and-robustness} shows that a population
with a very large initial variation in robustness evolved an increased ability
to absorb mutations after $T=400$ generations of stabilizing selection.
However, it is not very clear what kind of a reorganization in the
state spaces of these networks occur during the evolution. We measured the changes in
several system parameters to answer this question.

\subsection{Change in Dynamical Character }

\begin{figure}
\centering
\includegraphics[width=0.4\columnwidth]{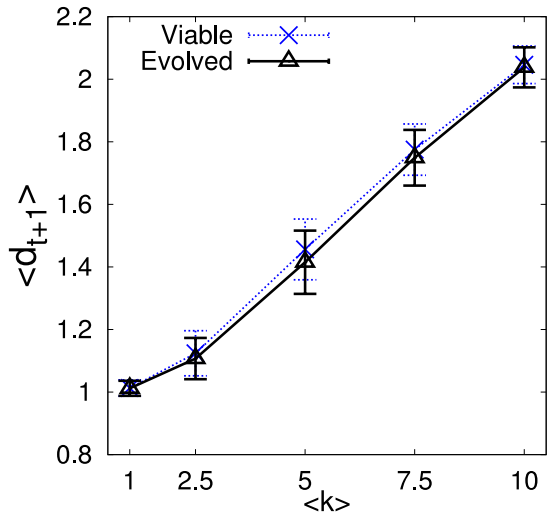}

\caption{Mean damage spreading, $\langle d_{t+1} \rangle$, after a one-bit perturbation
before (viable) and after evolution (evolved) for networks with $N=10$, measured 
exhaustively for all possible state pairs, averaged 
over 10000 samples each. The error bars represent one standard deviation (not standard error). Evolved networks (triangles) are slightly more ordered compared to
their viable ancestors. (The differences are statistically significant.)  $\langle d_{t+1} \rangle$ is always larger than unity due to our specific update rule. See the discussion in the text for details. The lines connecting the symbols are guides to the eye. 
(Probability distributions for $\langle d_{t+1} \rangle$ for certain values of  $\meank$  are given 
in supplementary Figs. S3 and S4.) 
\label{fig:matrixdistr-and-damage}}

\end{figure}

It has been analytically shown 
that the RTNs undergo a phase transition from order to chaos with increasing $\meank$  \citep{Rohlf:2002}. Obviously, 
the state space of an RTN is finite. Therefore, the expression states have to display
periodicity in development after at most $2^{N}$ steps. Thus, chaos here
is not a long-term aperiodic behavior; rather it corresponds to a dynamical regime in
which small perturbations percolate through the gene network 
\citep{AldanaReview,Luque:2000,Luque:2001}.
We quantify the dynamical character of a network  by comparing the time development
of two configurations, $\mathbf{s}(t)$ and $\mathbf{s}'(t)$,
that differ by one gene: we measure the mean number of different genes in time step $t+1$, $\langle d_{t+1} \rangle$, averaged over all possible $\mathbf{s}(t)$ and $\mathbf{s}'(t)$ pairs with ${\rm H}(\mathbf{s}(t),\mathbf{s}'(t))=1$. This is known as damage spreading or damage propagation\footnote{
This measure is used to determine the existence of a phase transition in RBNs and RTNs \citep{AldanaReview,DrosselRBNReview, Rohlf:2002}. However, it cannot be applied to our networks due to the way our update rule handles the genes without inputs. In our model, networks with low connectivity contain many nodes with no inputs, which retain the initial perturbation. The retained perturbation causes overestimation of the damage spreading (always above unity), concealing the phase transition as seen in Fig.~\ref{fig:matrixdistr-and-damage}. This incompatibility vanishes for larger $\meank$ as this detail in the update rule does not have any significance when almost all nodes have inputs. Nevertheless, we use the term ``damage spreading''   instead of ``damage-spreading rate'' when referring to $d_{t+1}$ to avoid a confusion.  
}
\citep{Luque:2000,Luque:2001,Derrida:1986}.
The dynamical behavior of random networks can
be quantified analytically using damage-spreading analysis \citep{Rohlf:2002, Luque:2001}. However, this analytical approach may not be applied to the \emph{evolved} networks 
\footnote{Obviously, all evolved networks satisfy the viability condition. 
For brevity, we use the term "evolved" instead of "evolved-viable" to specify these networks. Similarly, viable networks are essentially random as 
they are just a subset of the random networks, but we do not 
call them "random-viable" for the same reason.} 
as the selection process can tune a network to behave dynamically very differently than its ancestor. Therefore, we calculated $\langle d_{t+1} \rangle$
numerically for both viable and evolved networks, averaging ensembles of 10000 networks exhaustively over all possible configuration pairs, $\mathbf{s}(t)$ and $\mathbf{s}'(t)$,  for each $\meank$. As seen in Fig.~\ref{fig:matrixdistr-and-damage}, the $\langle d_{t+1} \rangle$ increases monotonically with increasing connectivity,
indicating that highly connected networks are more chaotic on average.
The evolved networks are slightly more ordered (on average) than their
viable ancestors. These results indicate that the dynamical behavior of the evolved
networks is not much different than that of the viable networks from
which they are descended.

\begin{figure}
\includegraphics[clip,width=0.4\linewidth]{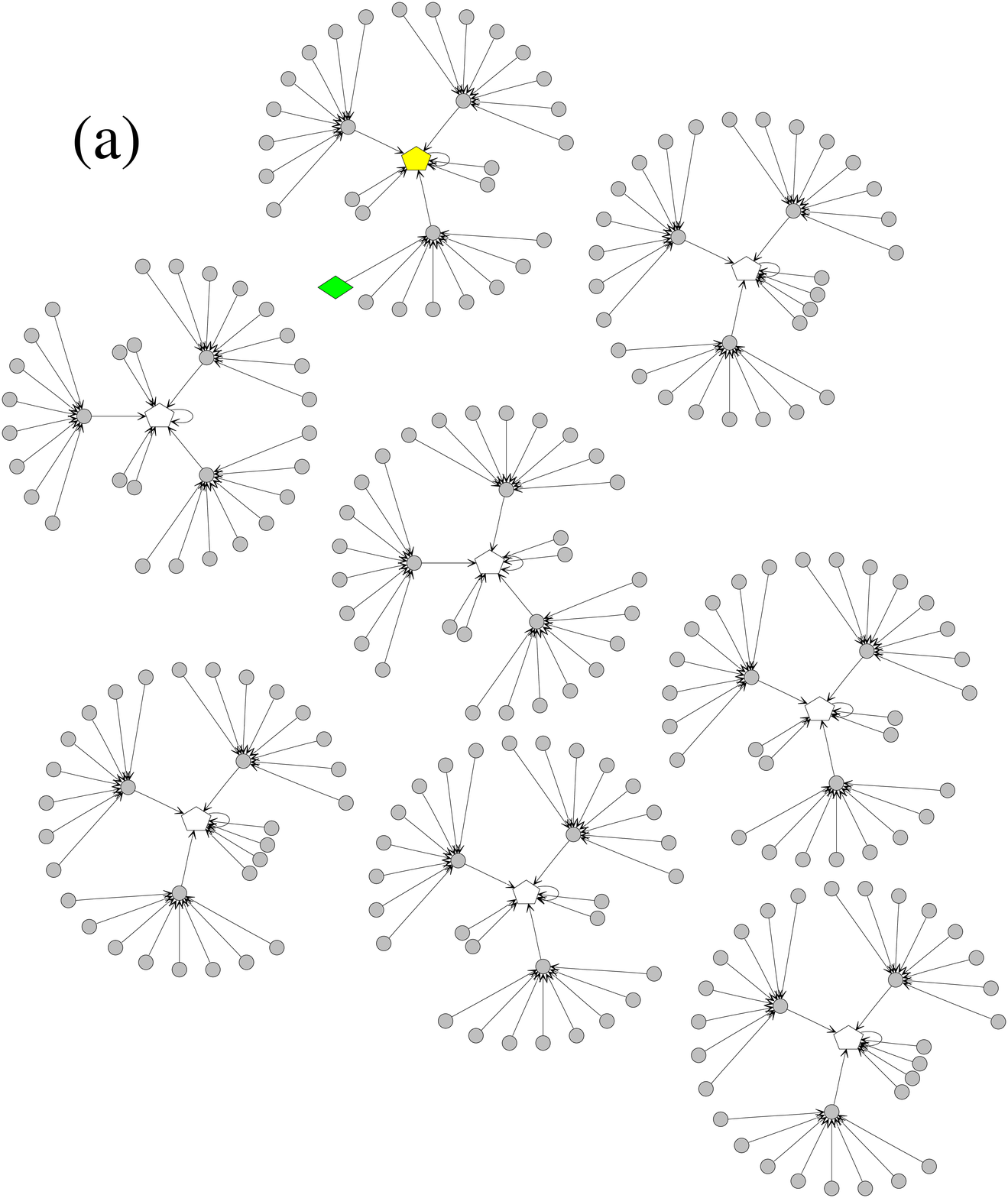}\includegraphics[clip,width=0.55\linewidth]{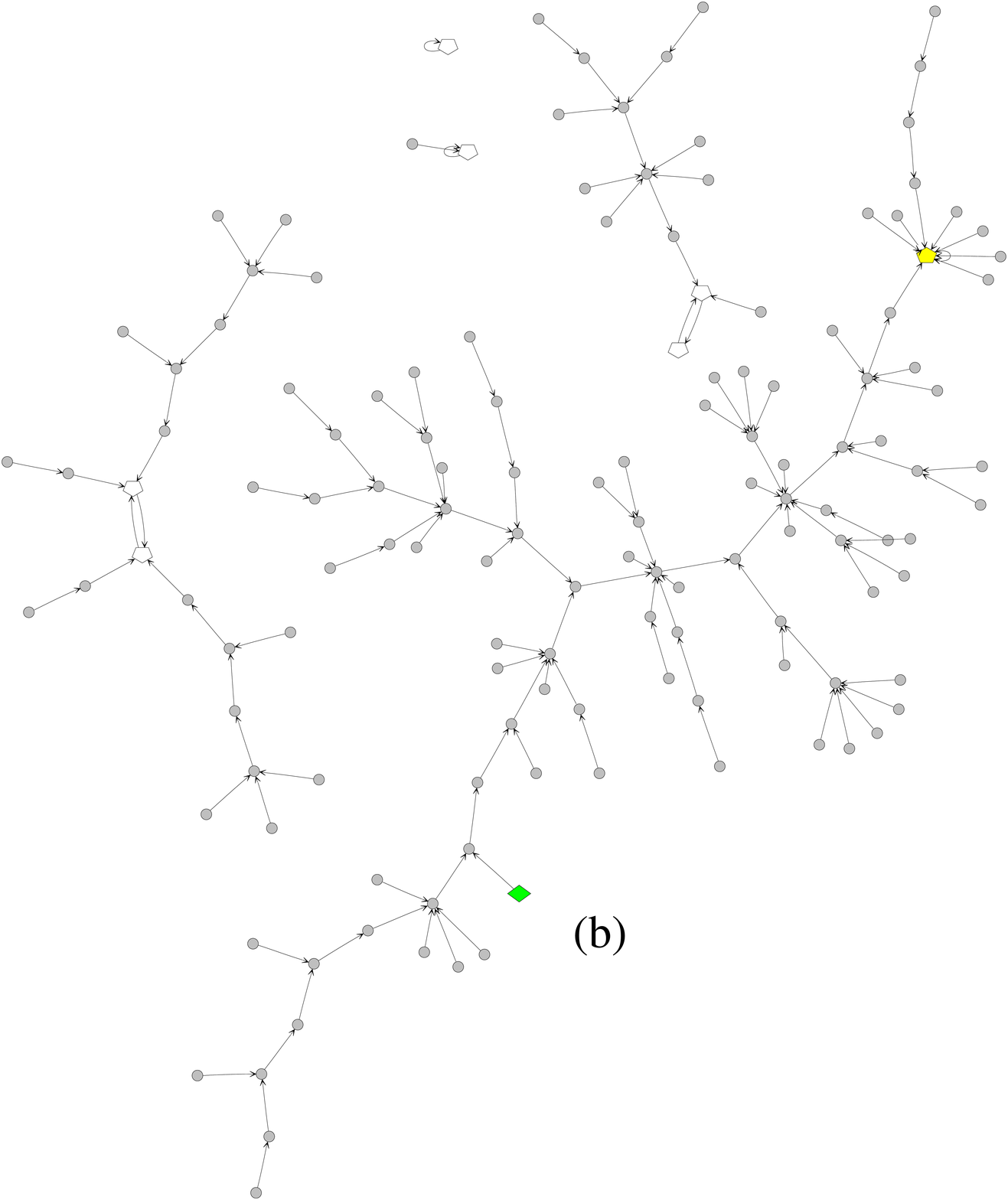}

\caption{Two typical state-space structures shown as directed graphs. Each node
represents a gene expression state, and the links represent developmental transitions
from $t$ to $t+1$. Arrows indicate the direction of flow. Each state
drains into an attractor (pentagons) through transient states. (a)
State space of a sample network with $N=8$ and $\langle k\rangle=1$.
There are eight basins, each having a fixed-point attractor. The principal
basin (at the top) contains the initial (diamond shaped node) and
final states. Note that the basins are quite symmetric, and transients
are very short as the system behaves less chaotically when the connectivity
of the network is low. (b) Basin of a sample network with $N=8$ and
$\langle k\rangle=8$. The principal basin occupies a large portion
of the state space. The basin on the left crosses the symmetry plane
with a 2-cycle (note the symmetry of the branches). The other basins
have mirror images on the other side of the state space (not shown).
The principal basin is significantly larger than the others. The high
connectivity makes the network behave more chaotic, creating basins
with broadly distributed sizes and branch lengths, and significantly
longer transients. See text for details. Generated using Graphviz
\citep{graphviz}. \label{fig:State-space-layout}}
\end{figure}

\subsection{Change in the State-Space Structure}

\begin{figure}
\includegraphics[clip,width=0.33\columnwidth]{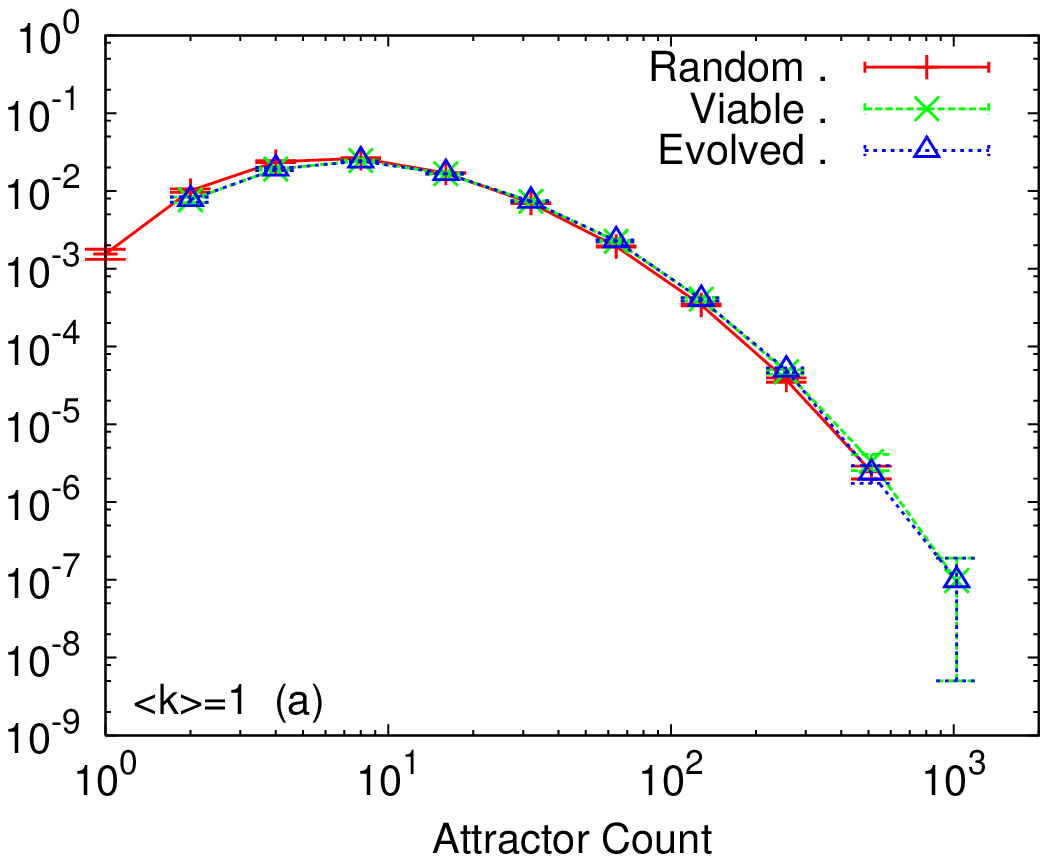}\includegraphics[clip,width=0.33\columnwidth]{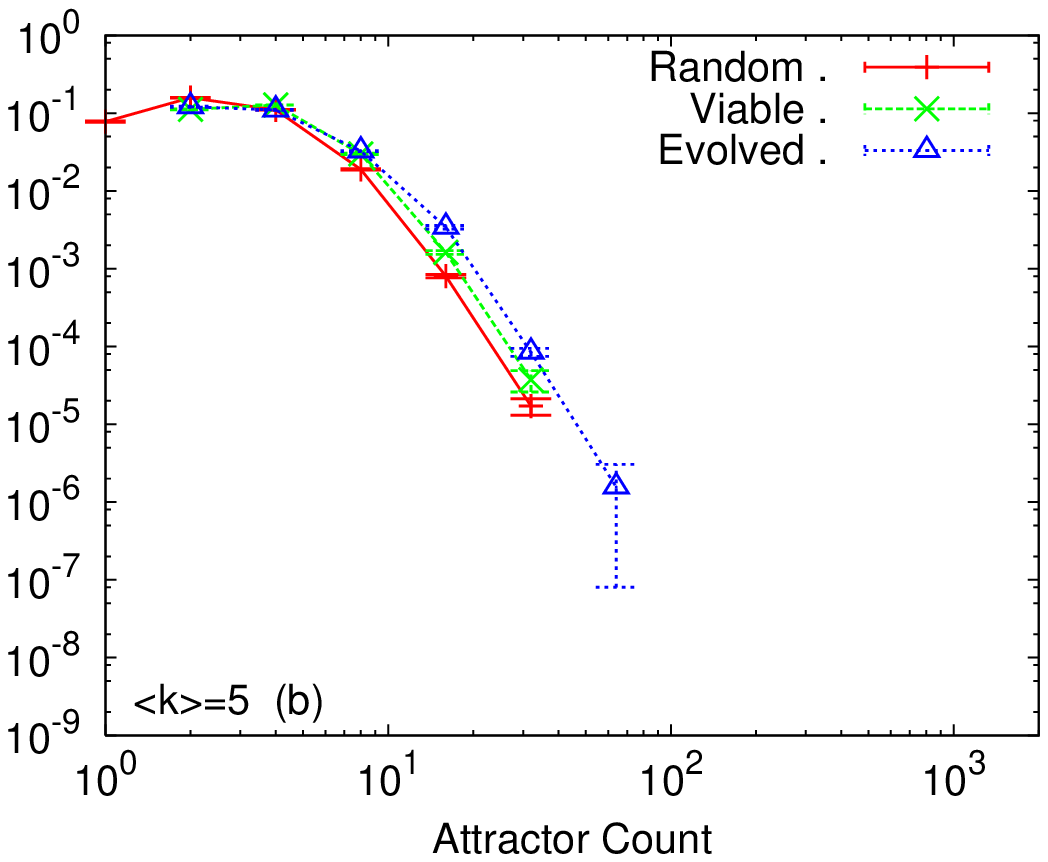}\includegraphics[clip,width=0.33\columnwidth]{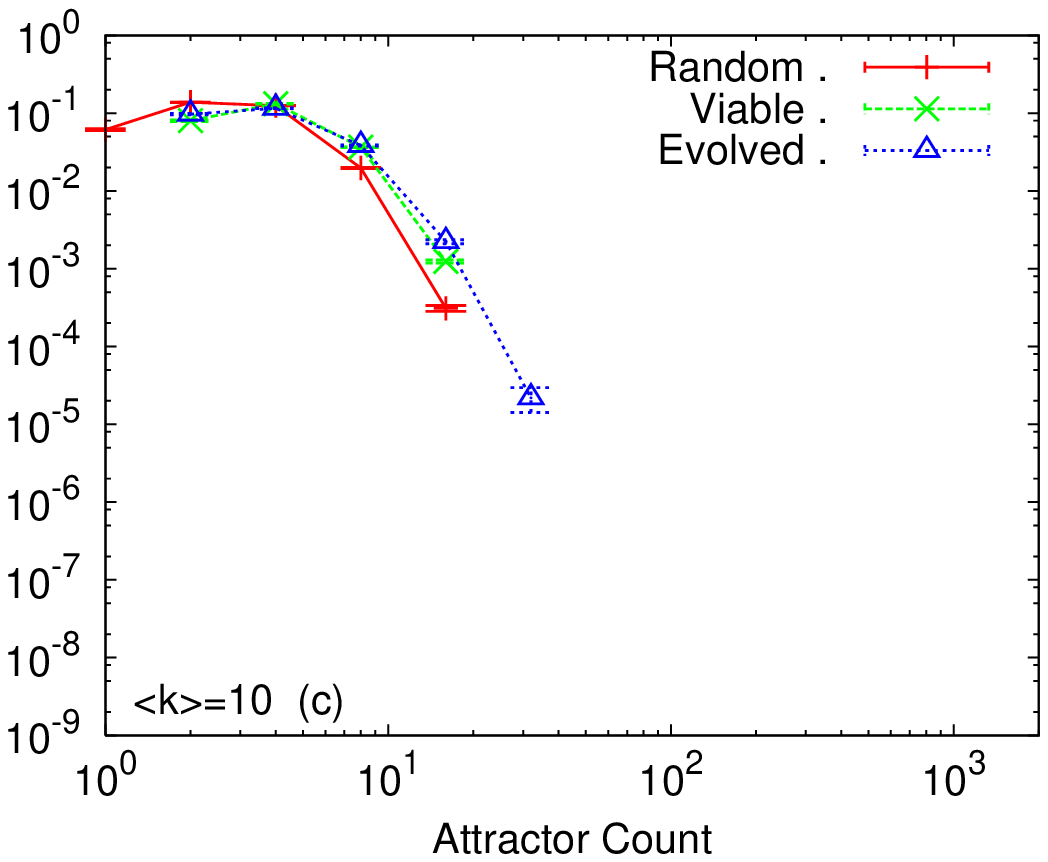}

\includegraphics[clip,width=0.33\columnwidth]{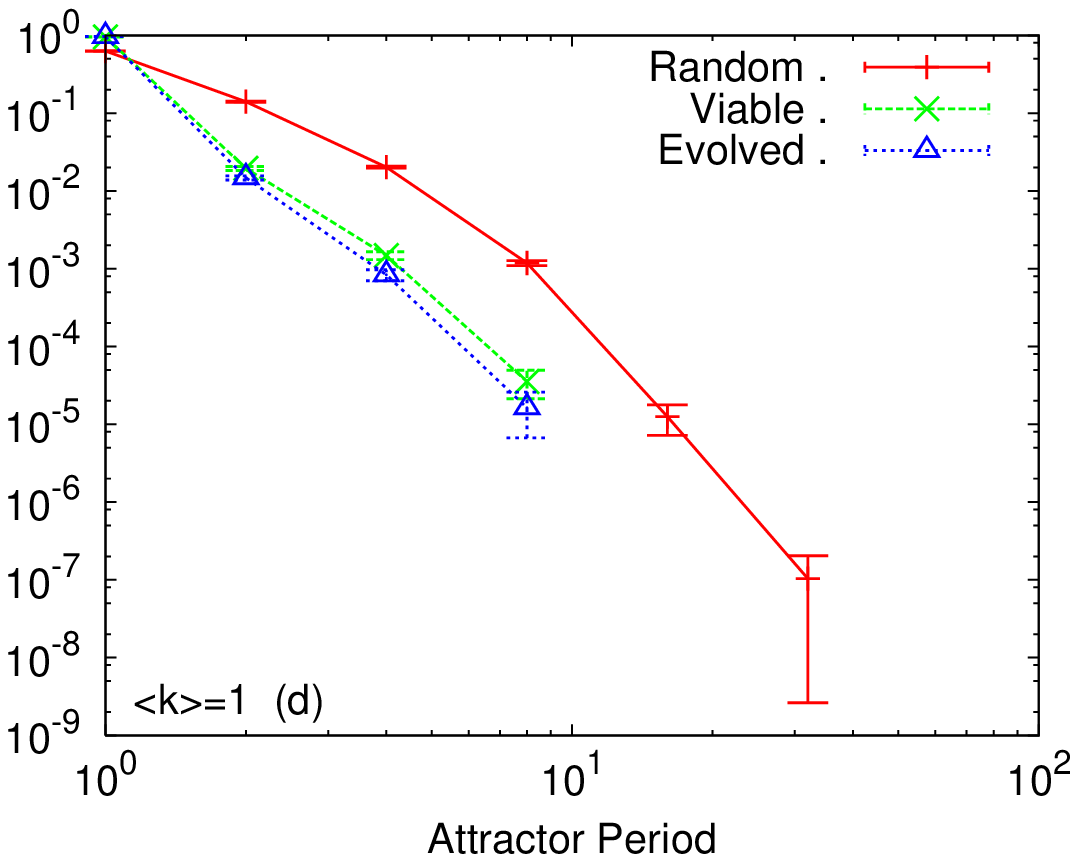}\includegraphics[clip,width=0.33\columnwidth]{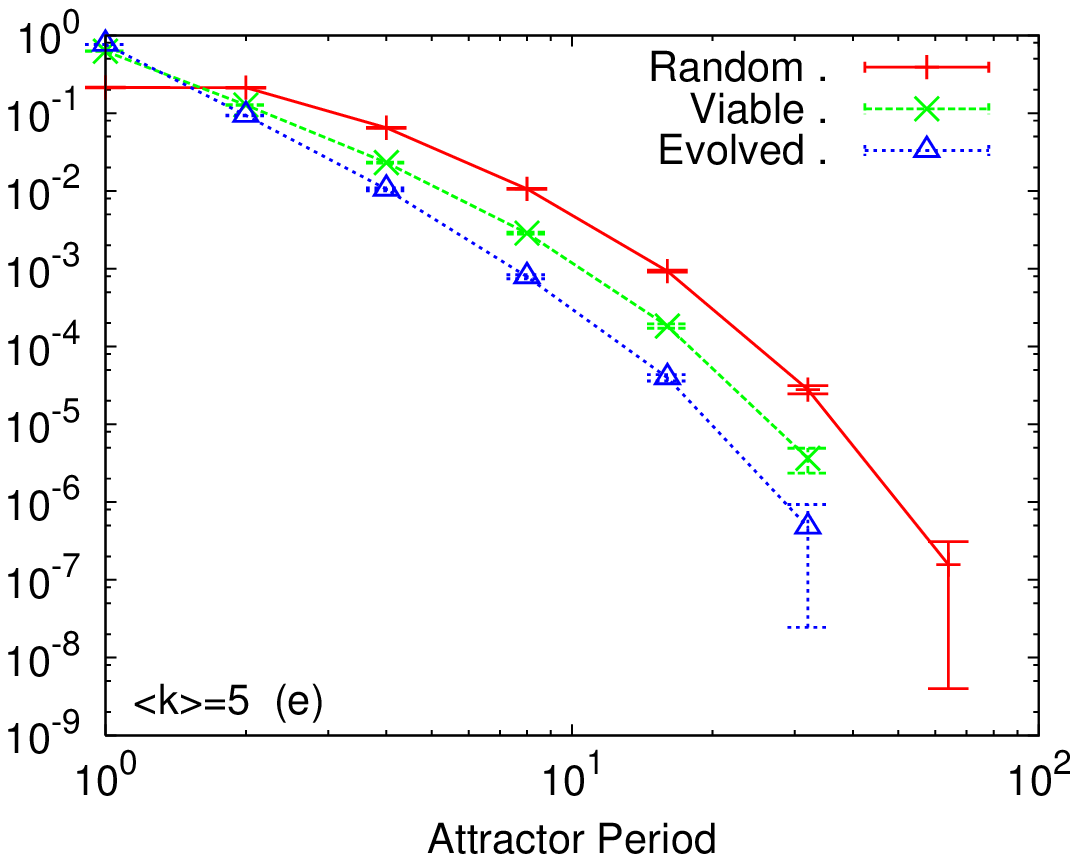}\includegraphics[clip,width=0.33\columnwidth]{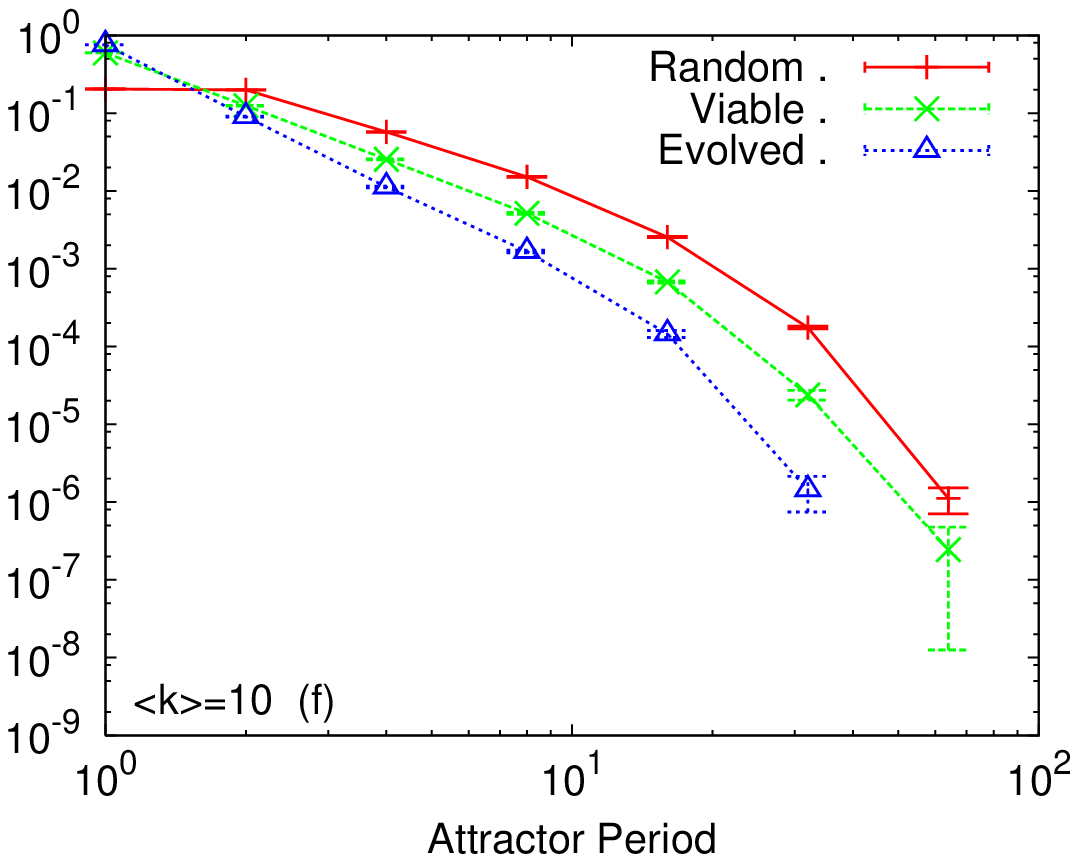}

\includegraphics[clip,width=0.33\columnwidth]{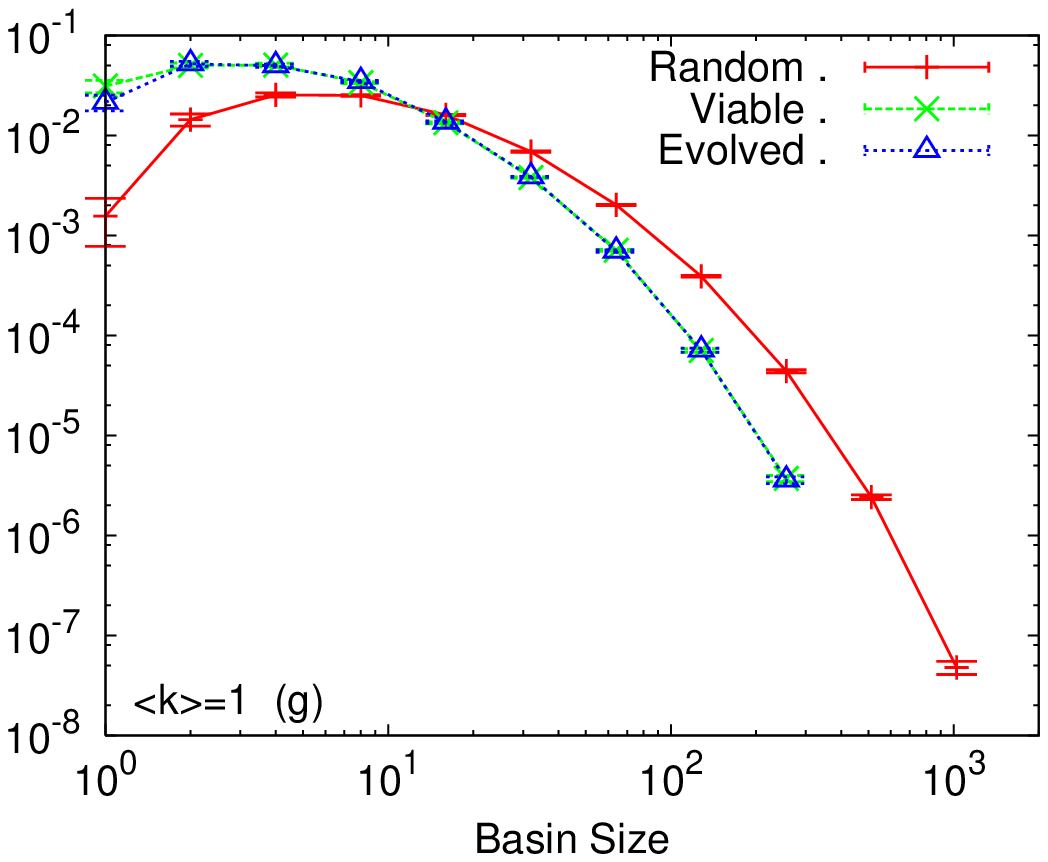}\includegraphics[clip,width=0.33\columnwidth]{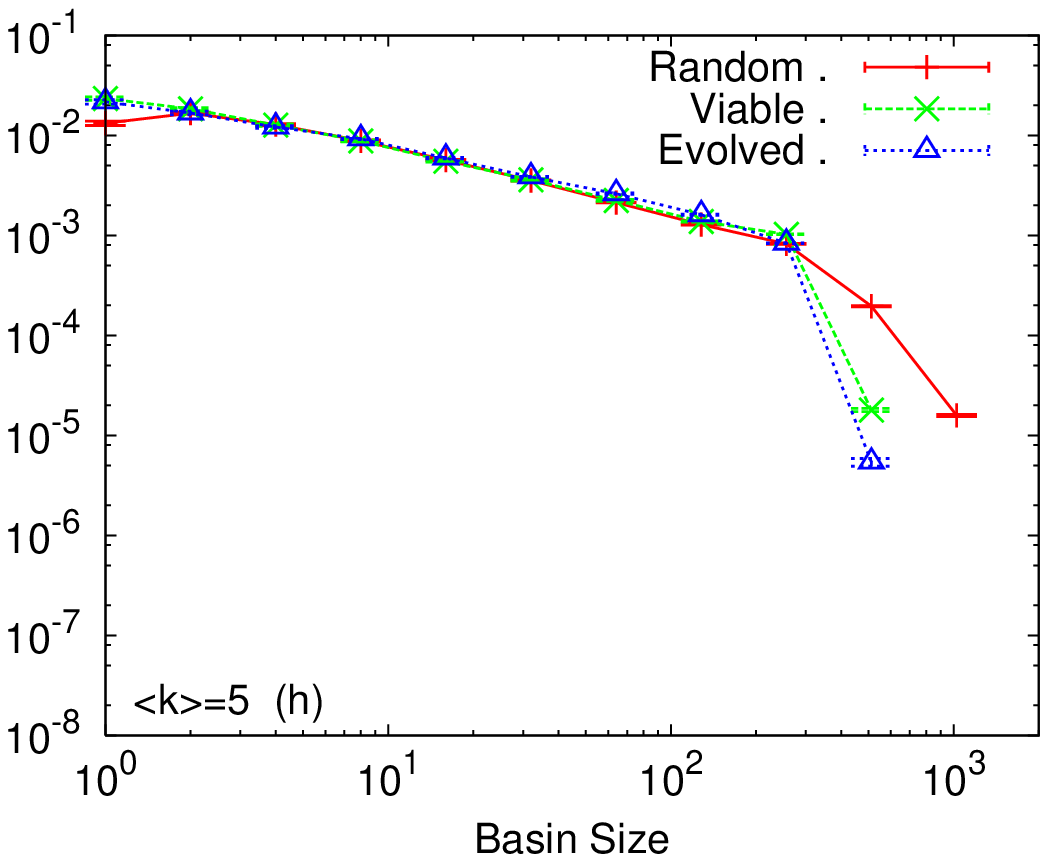}\includegraphics[clip,width=0.33\columnwidth]{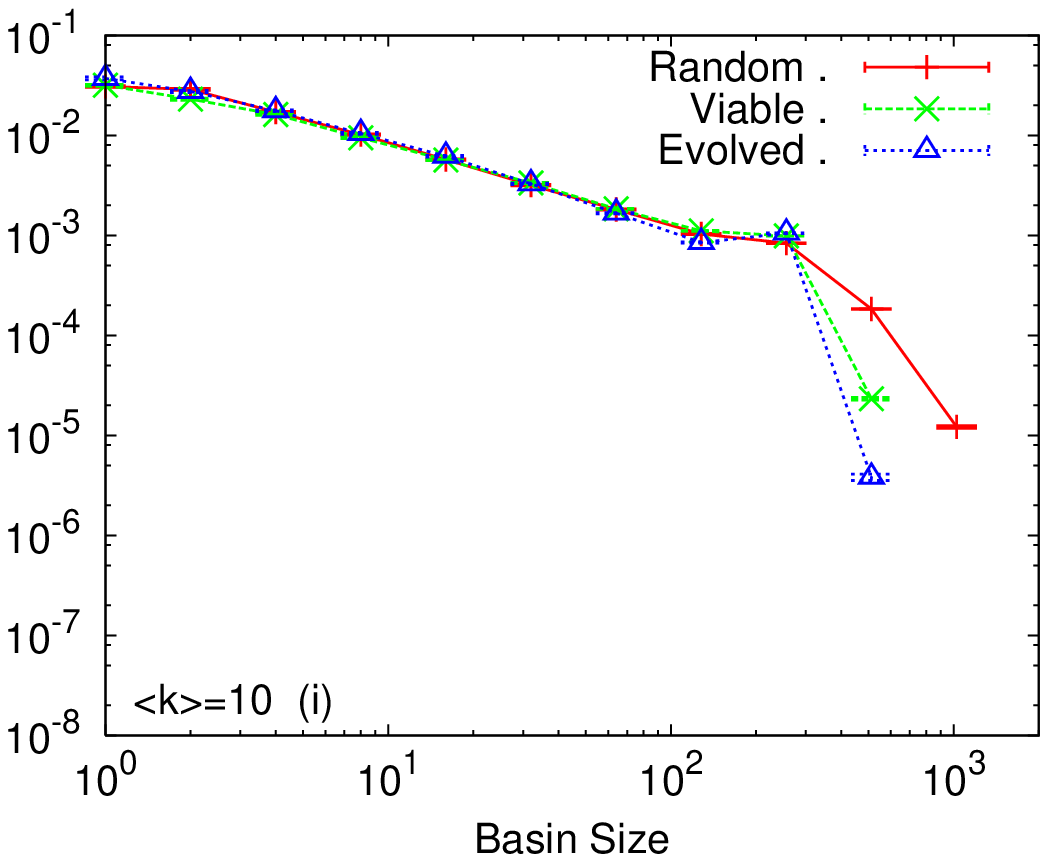}

\caption{Probability distributions for the attractor-count (first row), attractor-period
(second row), and basin-size (all basins, third row) distributions
for $N=10$ and $\langle k\rangle=1,$ 5, and 10 (columns 1, 2, and
3, respectively). Each curve represents an average over 20,000 realizations
for the {}``random'' networks, and 10,000 realizations for the {}``viable''
and {}``evolved'' ones. For the evolved networks, we did not use
clan averages to avoid a bias. Instead, we picked one sample network
from each evolved clan. Error bars were calculated by grouping the
data: Each data set was divided into groups of 1000 samples and the
average distribution for each group was calculated. Then, the error
calculations were performed on the new set of averaged distributions.
The data plotted on log-log scale were histogrammed using exponential
bins (0, 1, 2-3, 4-7,...) to reduce the noise. 
(See supplementary Fig. S1 for histograms with linear bins.) 
The basin size does
not include the size (period) of the attractor. The evolved attractor-count
and attractor-period distributions are shifted toward those of viable networks 
with lower $\meank$, indicating that evolved networks display a slightly
more ordered character. The basin-size distributions for the evolved
networks seem to overlap with the ones for the viable networks except
for the largest basins, which become less probable after selection.
The lines connecting the symbols are guides to the eye.\label{fig:stats1}}
\end{figure}

\begin{figure}
\includegraphics[clip,width=0.33\columnwidth]{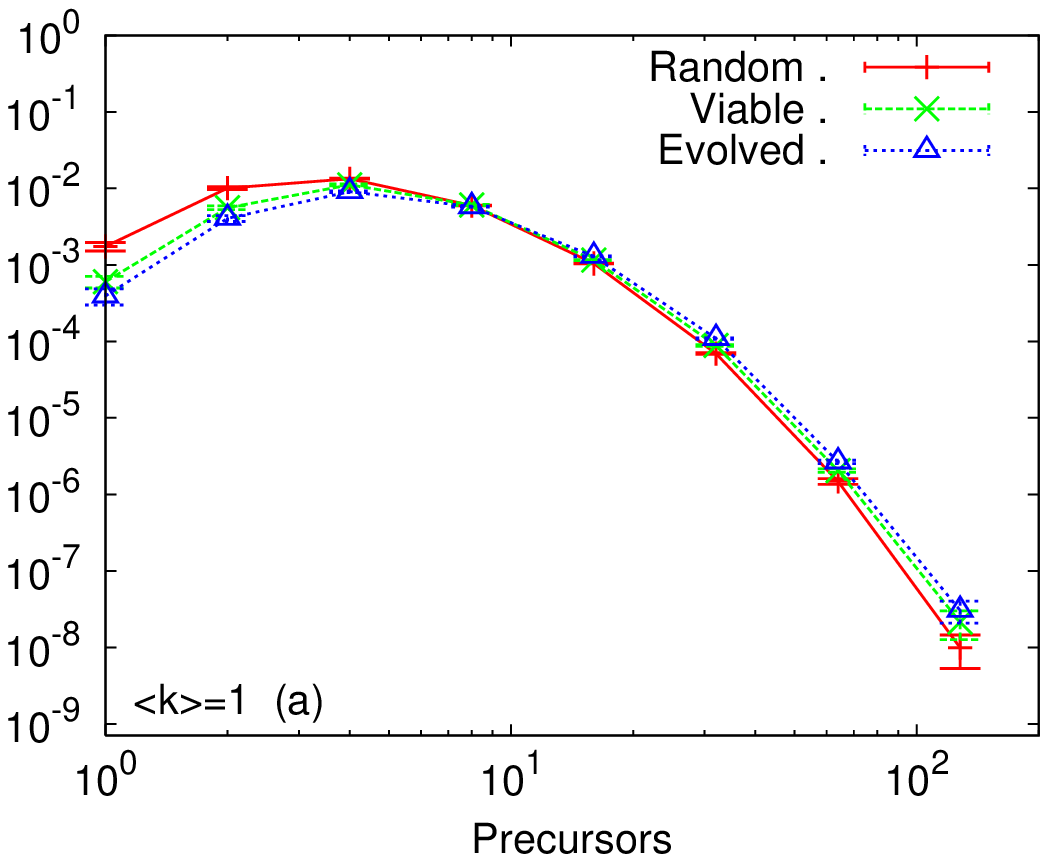}\includegraphics[clip,width=0.33\columnwidth]{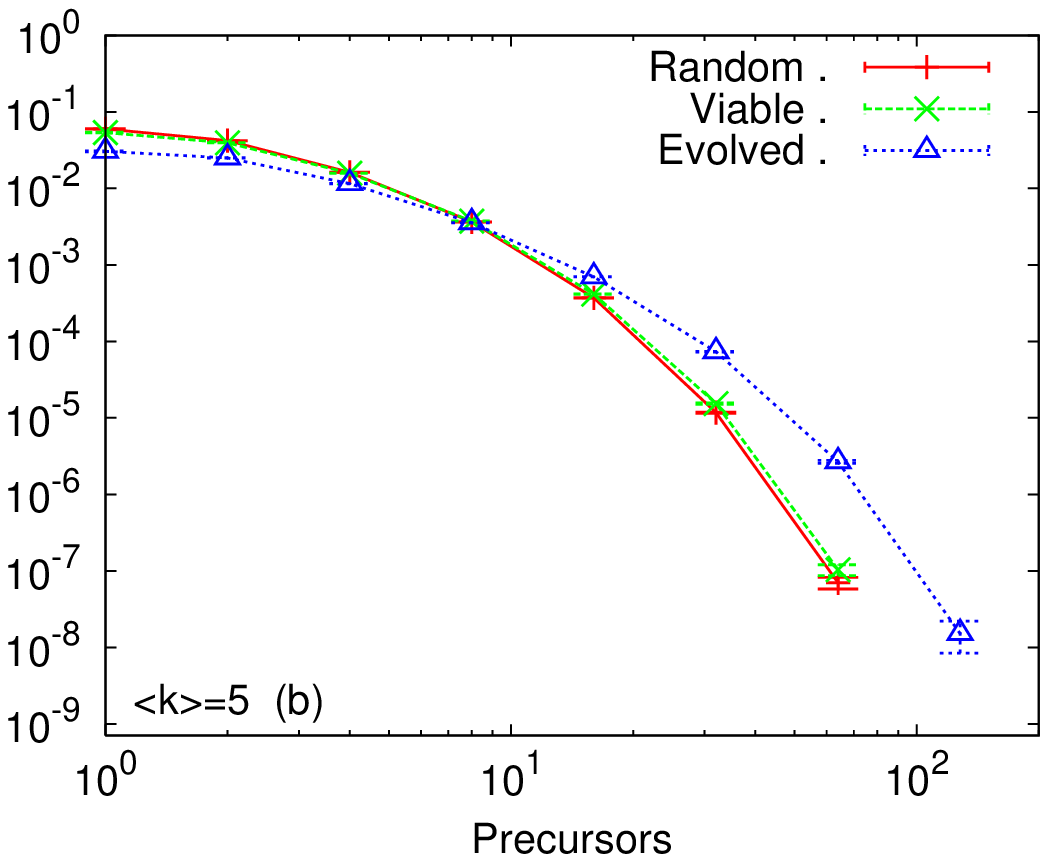}\includegraphics[clip,width=0.33\columnwidth]{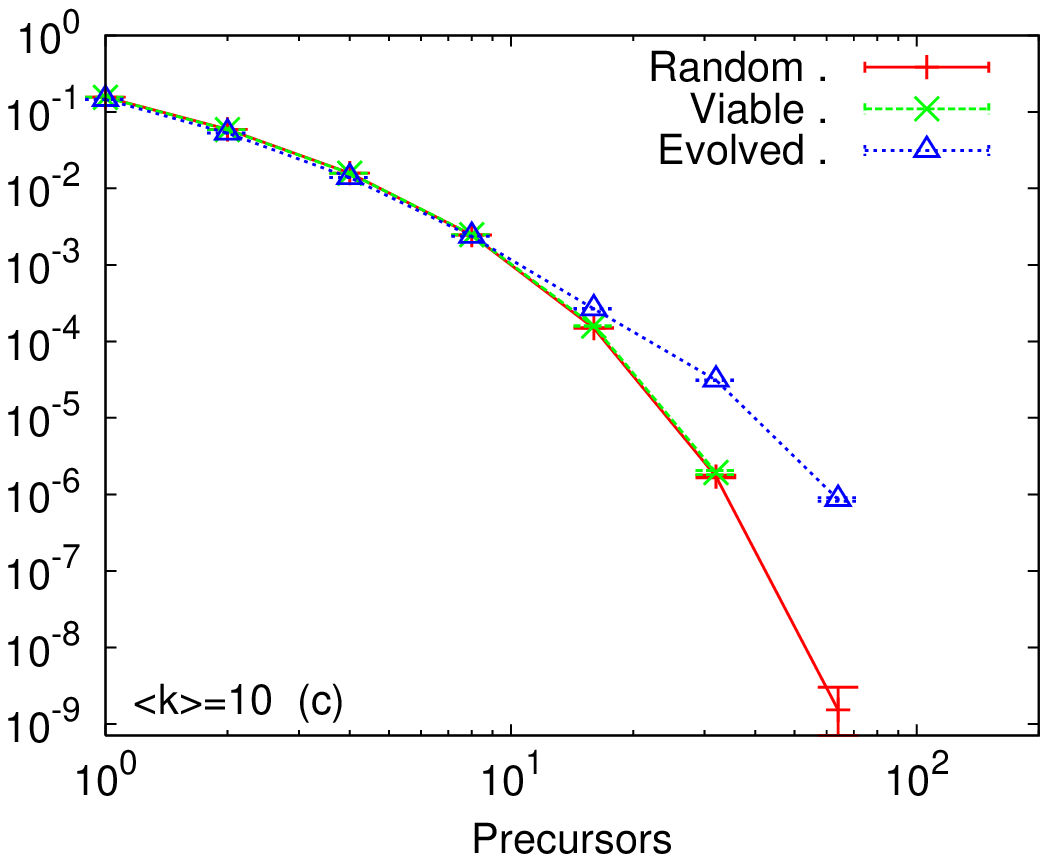}

\includegraphics[clip,width=0.33\columnwidth]{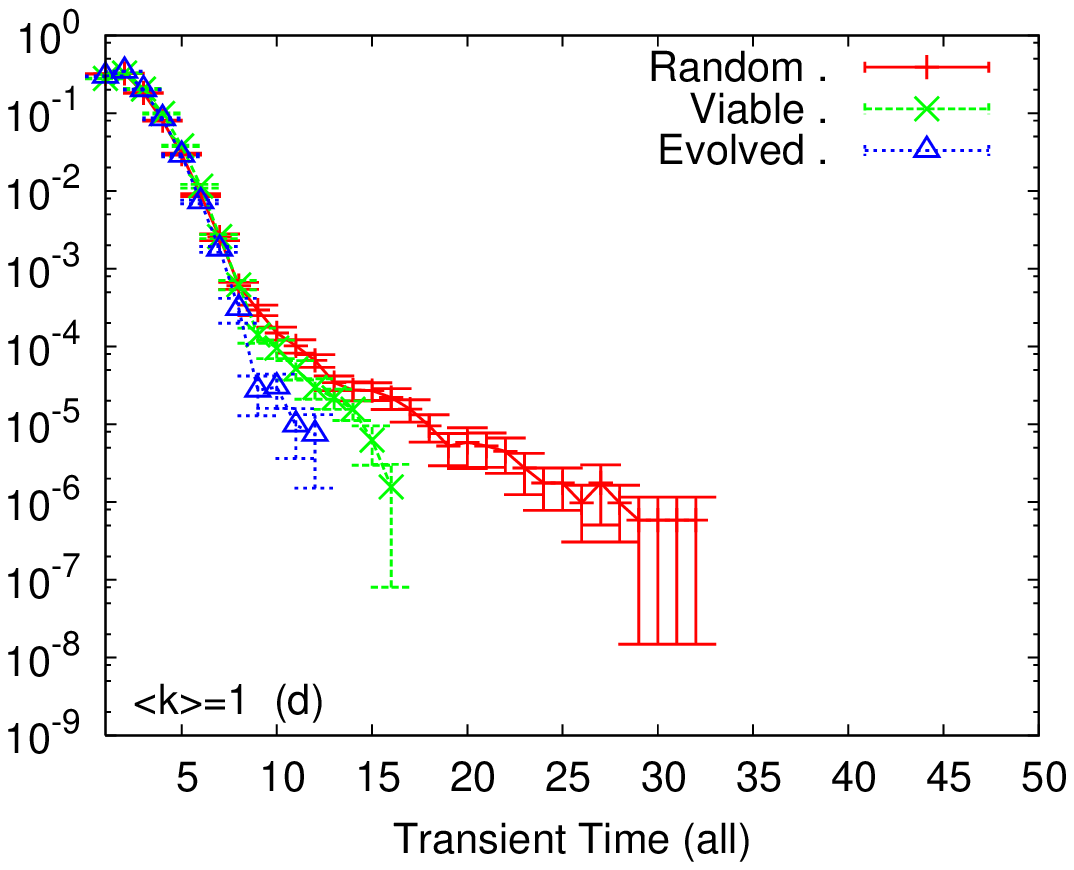}\includegraphics[clip,width=0.33\columnwidth]{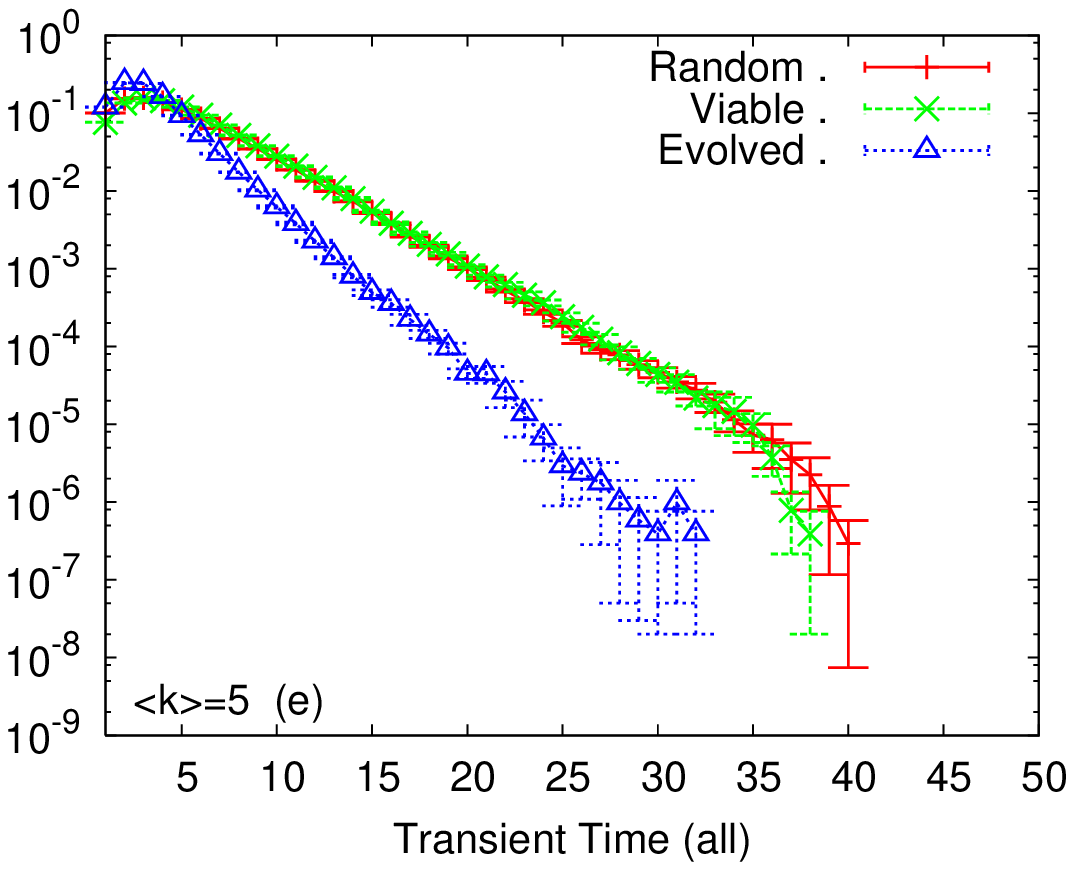}\includegraphics[clip,width=0.33\columnwidth]{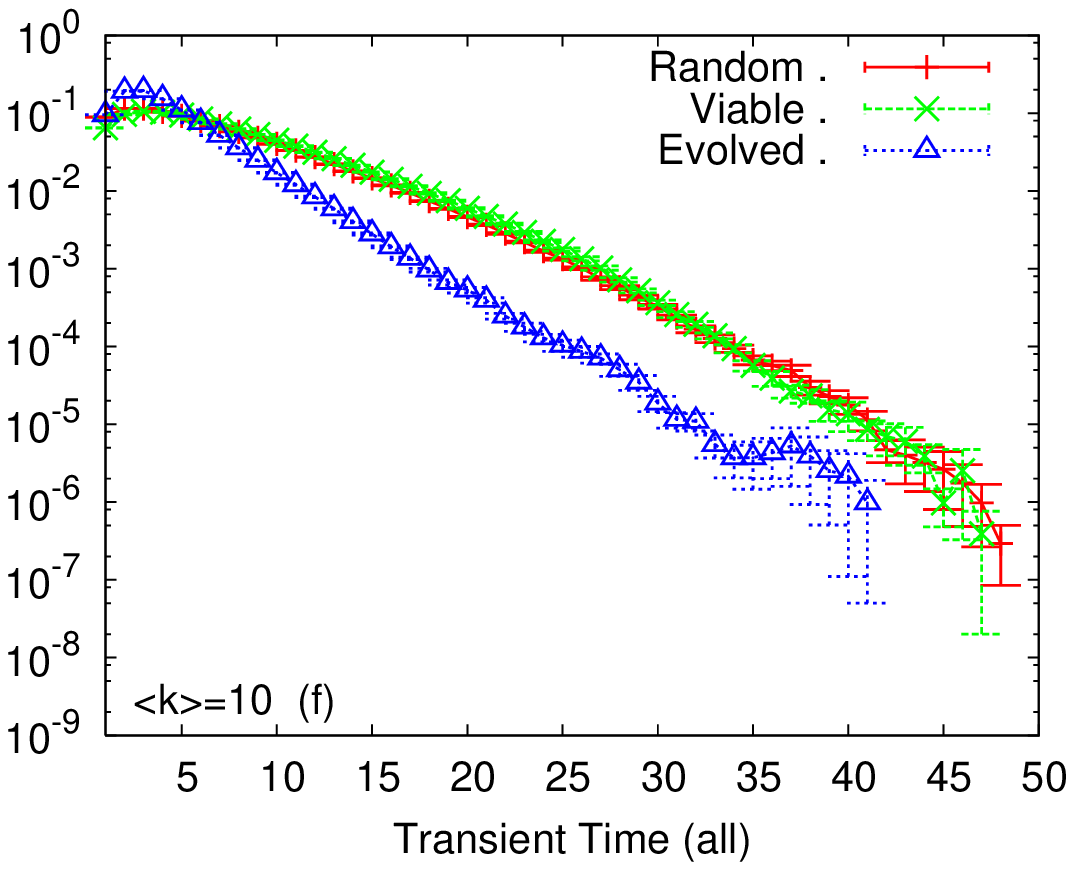}

\includegraphics[clip,width=0.33\columnwidth]{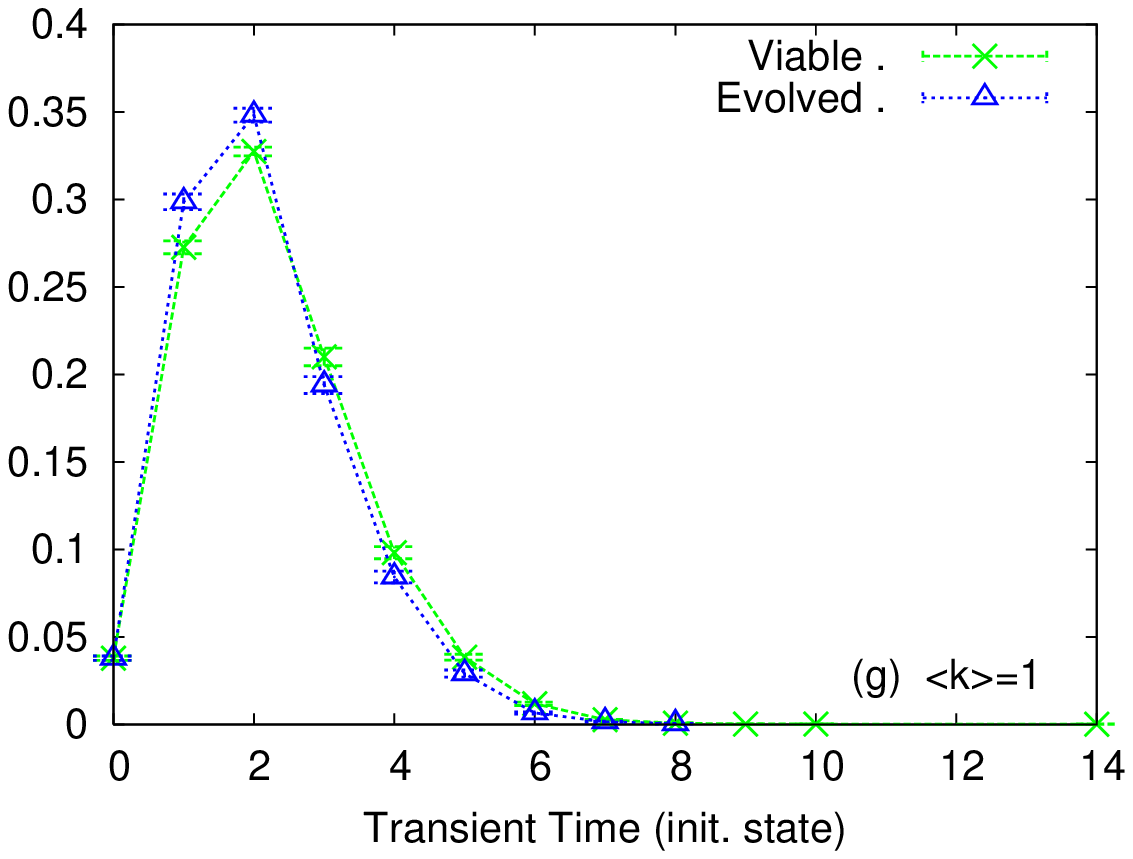}\includegraphics[clip,width=0.33\columnwidth]{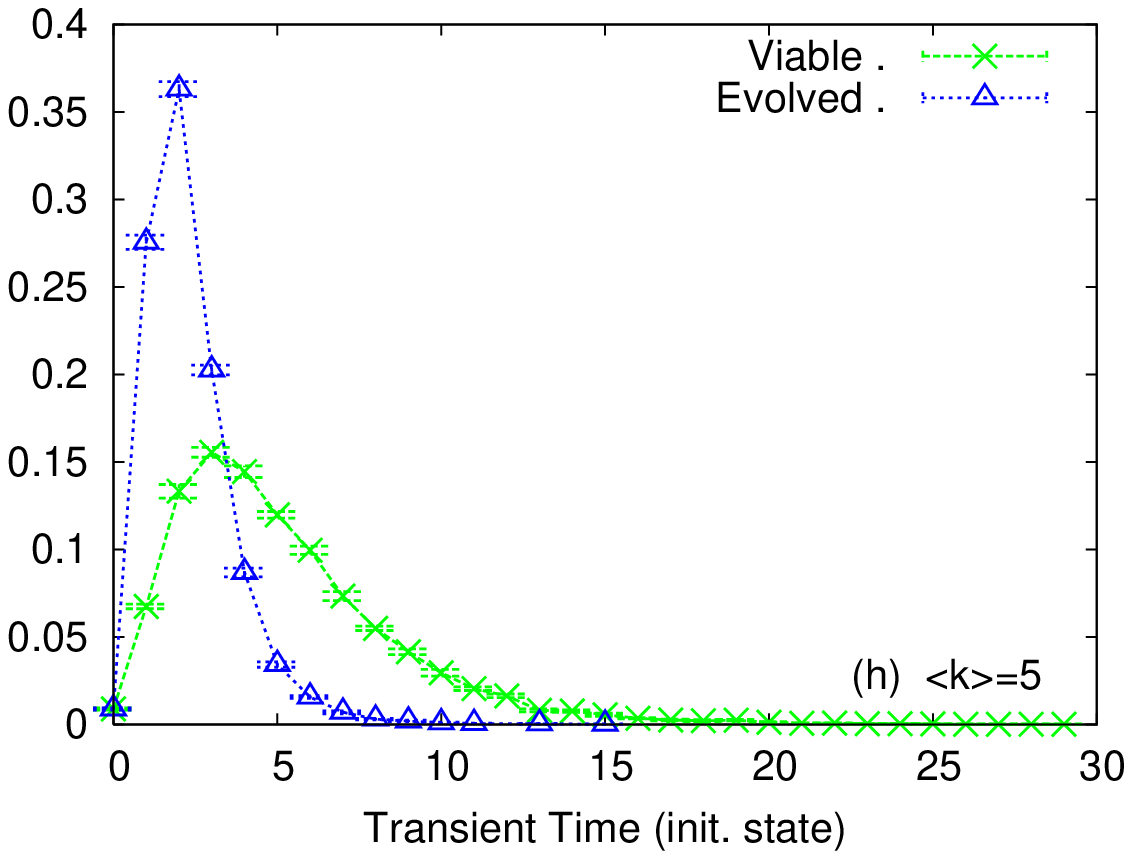}\includegraphics[clip,width=0.33\columnwidth]{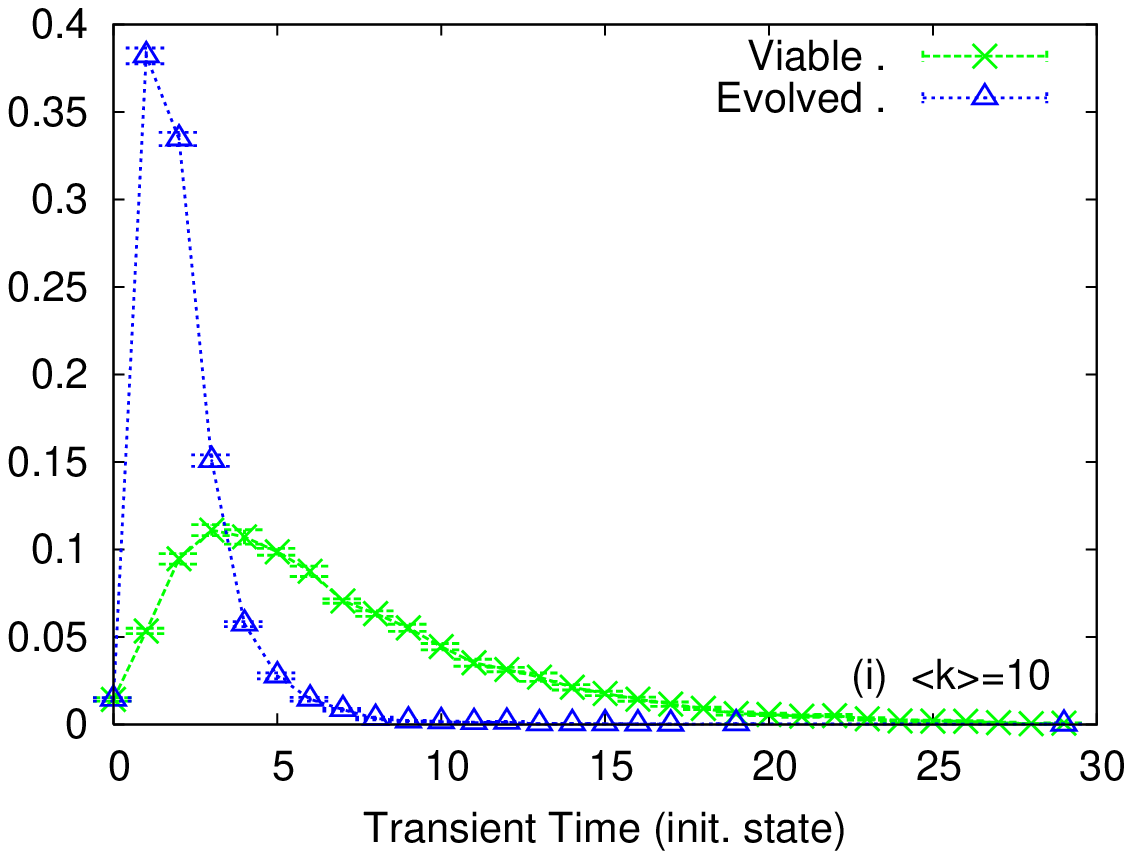}

\caption{Probability distributions for precursor (indegree of the state-space
graph) distribution (first row), transient time for all states (second
row), and transient time for the initial state (third row) for $N=10,$
$\langle k\rangle=1,$ 5, and 10 (columns 1, 2, and 3, respectively).
The second row shows the distributions for the lengths of the transients starting from all possible states. The third row, however, shows the distribution for $\tau$, the length of the transient starting from $\initst$, the original initial state. Again, changes in the distributions indicate that evolved networks display a slightly more ordered character. The data plotted on log-log scale were histogrammed using exponential bins (0, 1, 2-3, 4-7,...)
and the other plots were histogrammed using linear (0,1,2...) bins.
See the caption of Fig.~\ref{fig:stats1} for the details of the
data analysis. The lines connecting the symbols are guides to the
eye.\label{fig:stats2}}
\end{figure}

We also analyzed the state spaces of random, viable, and evolved networks
\citep{Socolar:2007}. Figures \ref{fig:stats1} and \ref{fig:stats2}
show these statistics for networks with $N=10$ and connectivities
$\meank=$1, 5, and 10. Due to the up-down symmetry of the system,
the state space is divided into two parts, where the dynamics on one
side is the mirror image of the other. Therefore,
the basin size of a fixed-point attractor cannot exceed half the size
of the state space, $\Omega/2=2^{N-1}$. Limit cycles, however, can
cross the symmetry plane that divides the state space (Fig.~\ref{fig:State-space-layout}(b)).
Therefore, their basins can contain up to $\Omega=2^{N}$ states. 

Typically, sparsely connected gene networks have many attractors (fixed
points and limit cycles) and, consequently, smaller basin sizes
on average, as depicted in Fig.~\ref{fig:State-space-layout}(a).
Increasing connectivity brings state spaces with fewer attractors
(Figs.~\ref{fig:stats1}(a), (b), and (c)), longer attractor (limit
cycle) periods (Figs.~\ref{fig:stats1}(d), (e), and (f)), and broadly
distributed basin sizes (Figs, \ref{fig:stats1}(g), (h), and (i)).
Densely connected networks can contain basins that occupy a large
portion, occasionally even all, of the state space. The basins of such networks
tend to have broadly distributed branch lengths as seen in Fig.~\ref{fig:State-space-layout}(b),
and their states also tend to have fewer precursors (Figures~\ref{fig:stats2}(a),
(b), and (c)). (All states, $\{\mathbf{s}(t)\}$, that go to state
$\mathbf{s}^{\dagger}$ at time step $t$+1 are precursors of $\mathbf{s}^{\dagger}$.)
Therefore, mean transient time (total number of steps from a state
to the attractor) on such networks are typically larger than on 
networks with lower connectivity (Figs.~\ref{fig:stats2}(d),
(e), and (f)). 

These changes in the state-space characteristics are consistent with the
damage-spreading measurements. The distributions for the evolved networks
are shifted toward the distributions of more ordered networks
of lower connectivity. This means evolved networks are slightly more ordered, as the damage-spreading measurements shown in Fig.~\ref{fig:matrixdistr-and-damage}(b)
suggest. For example, the distribution of the number of attractors
(Figs.~\ref{fig:stats1}(a), (b) and (c)) for the networks with low
connectivity ($\meank=1)$ has a longer tail compared to the networks
with $\meank=5$ and 10. Similarly, the distributions for the evolved
networks have slightly longer tails compared to those of viables of
the same connectivity. The same effect can be seen in the attractor-period
(Figs.~\ref{fig:stats1}(d), (e), and (f)), precursor (Figs.~\ref{fig:stats2}(a),
(b), and (c)), and transient-time distributions (Figs.~\ref{fig:stats2}(d),
(e), and (f)), as well. The most significant changes are seen in the
transient-time distributions, indicating that basins with relatively
shorter branches are preferred by selection. The basin-size distributions
of viable and evolved networks, however, do not display much difference
as they virtually overlap.

\subsection{Change in Gene Expression Trajectory and the State-Space Structure
Nearby}

\begin{figure}
\includegraphics[clip,width=0.23\columnwidth]{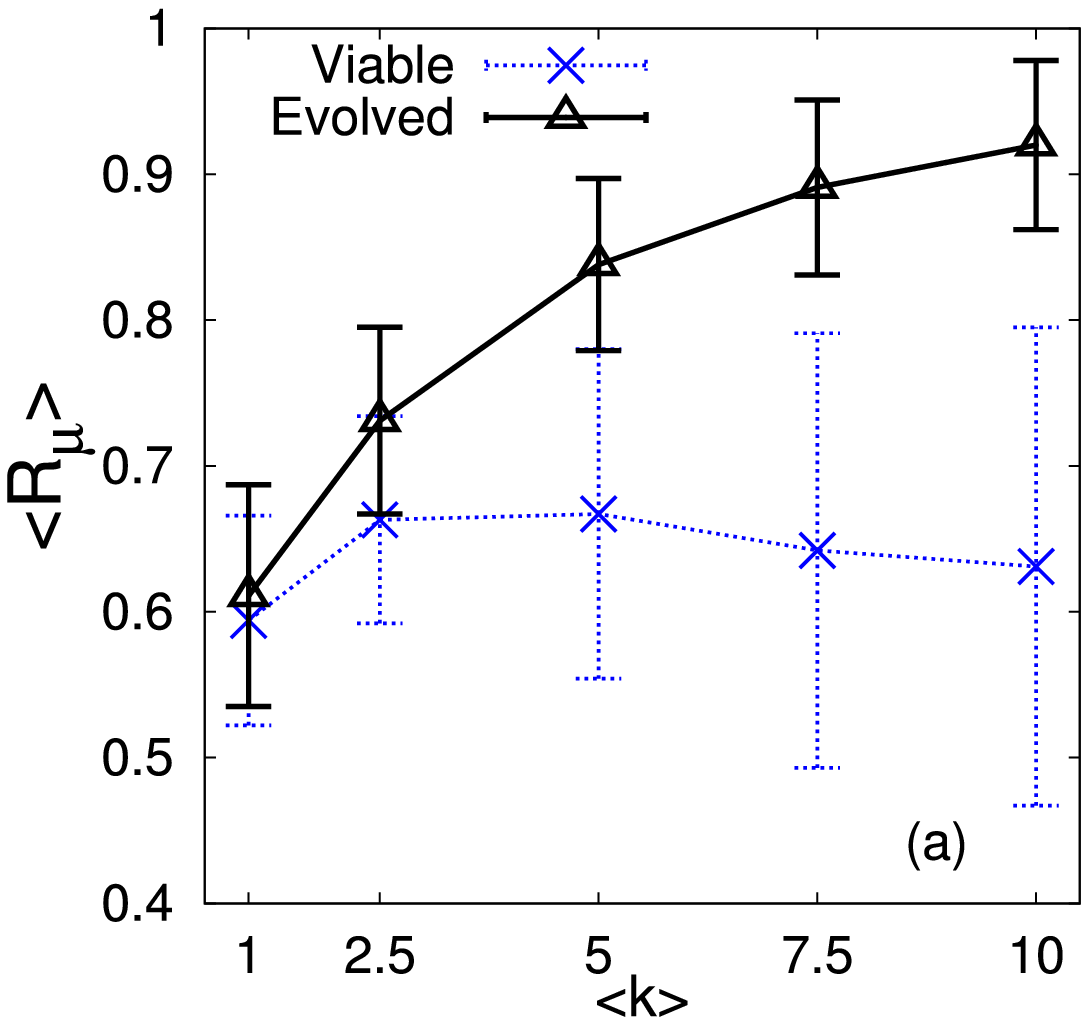}\hspace{.2truecm}\includegraphics[clip,width=0.22\columnwidth]{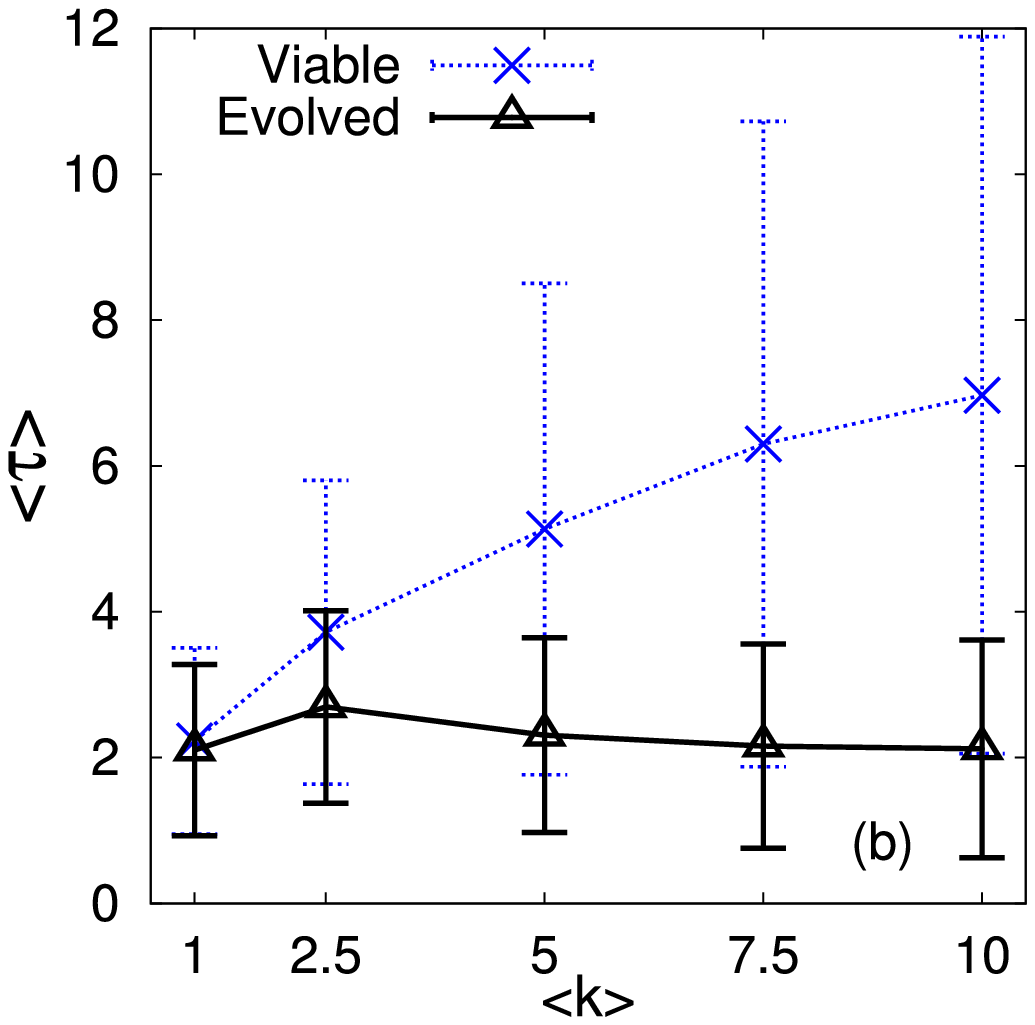}\hspace{.2truecm}\includegraphics[clip,width=0.23\columnwidth]{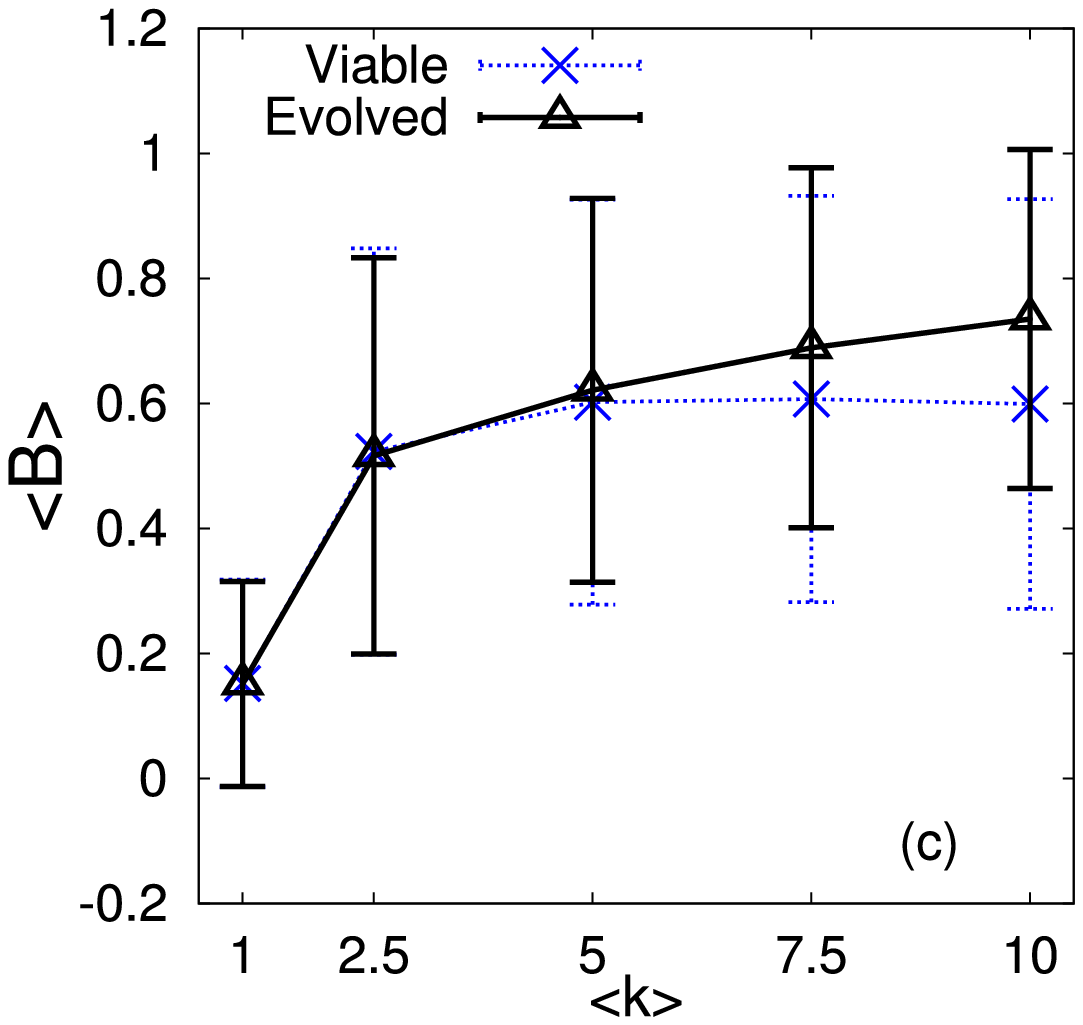}\hspace{.2truecm}\includegraphics[clip,width=0.23\columnwidth]{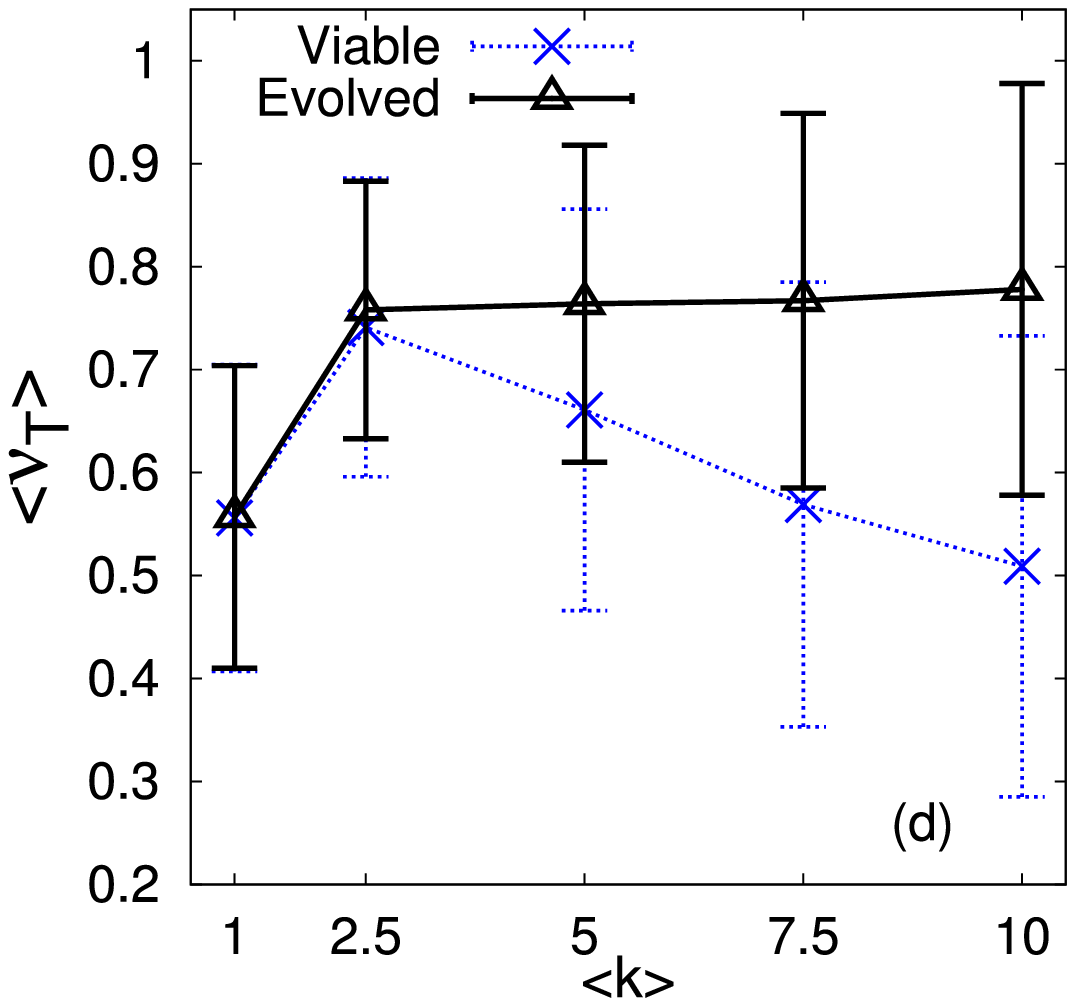}

\vspace{.2truecm}

\includegraphics[clip,width=0.23\columnwidth]{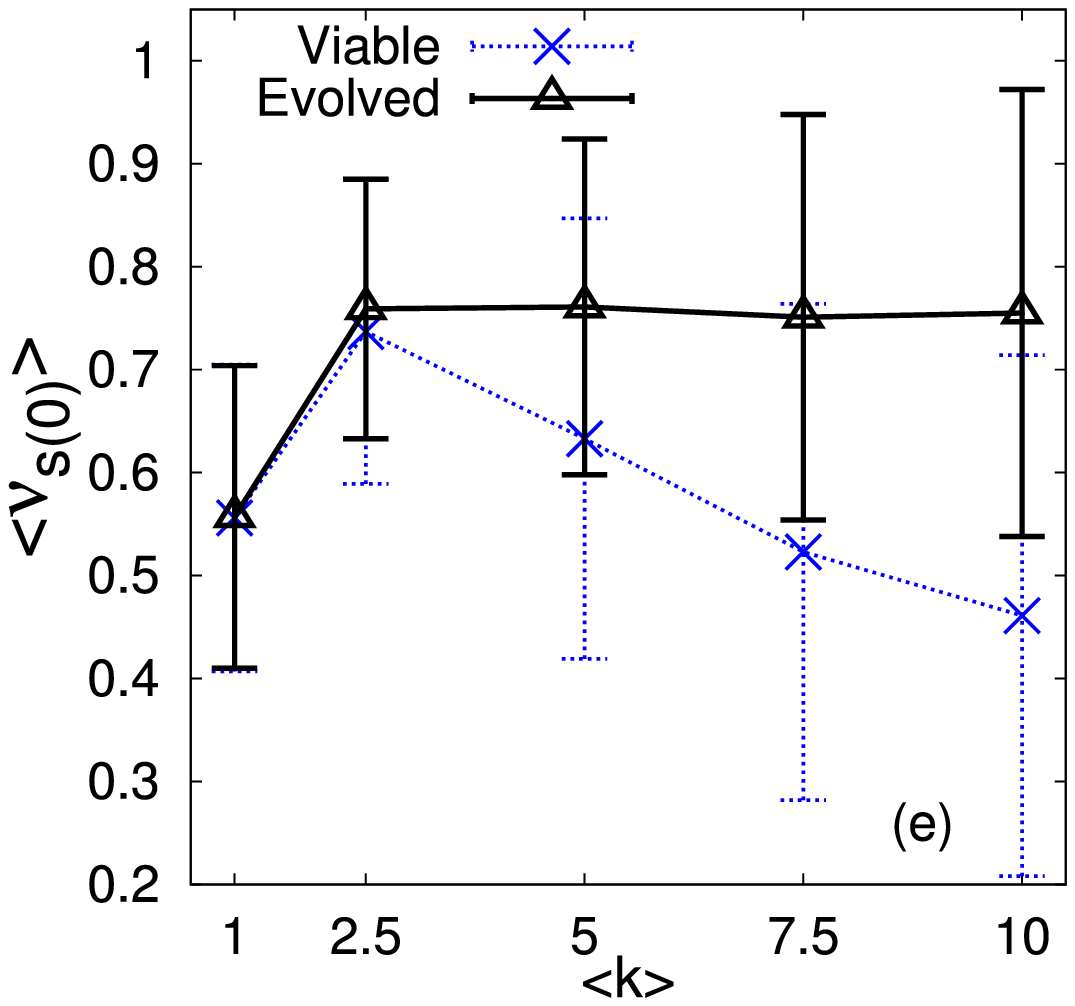}\hspace{.2truecm}\includegraphics[clip,width=0.23\columnwidth]{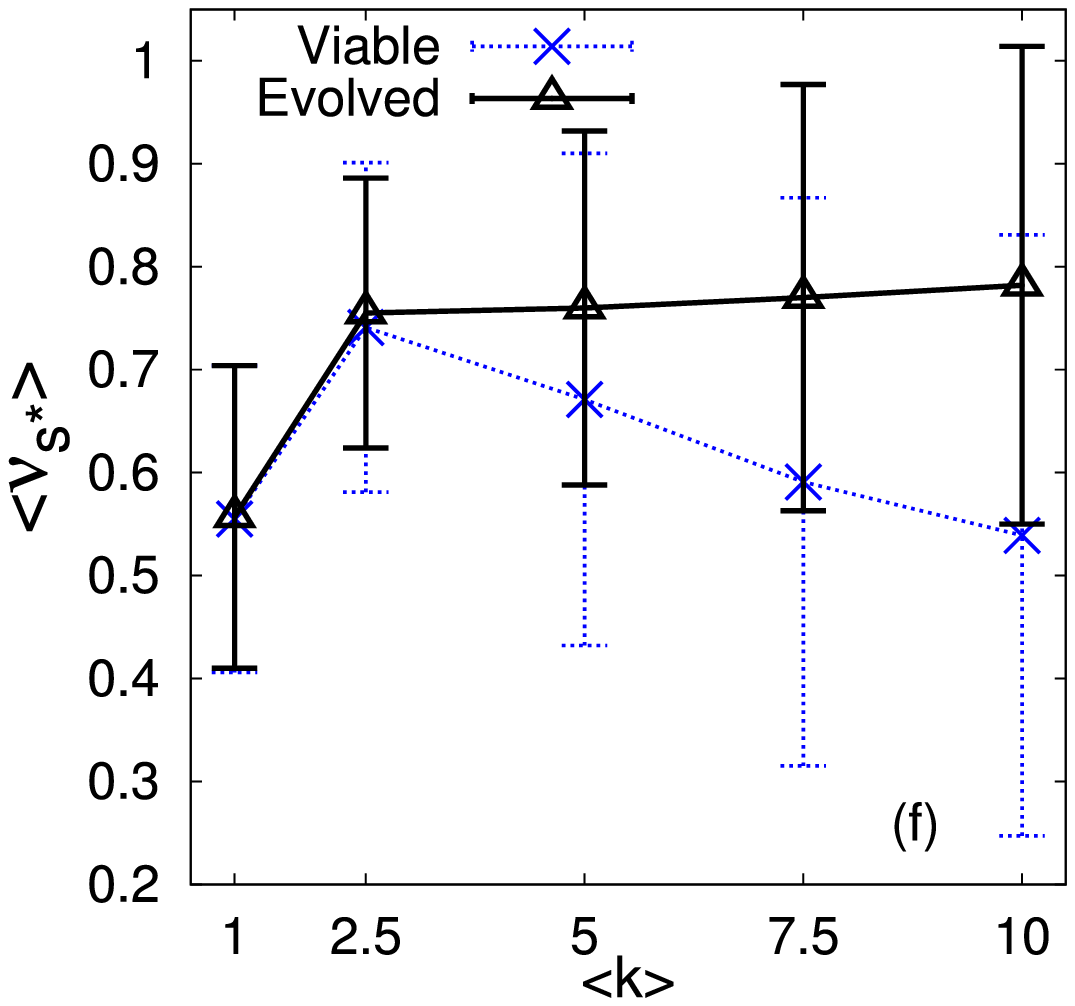}\hspace{.2truecm}\includegraphics[width=0.23\columnwidth]{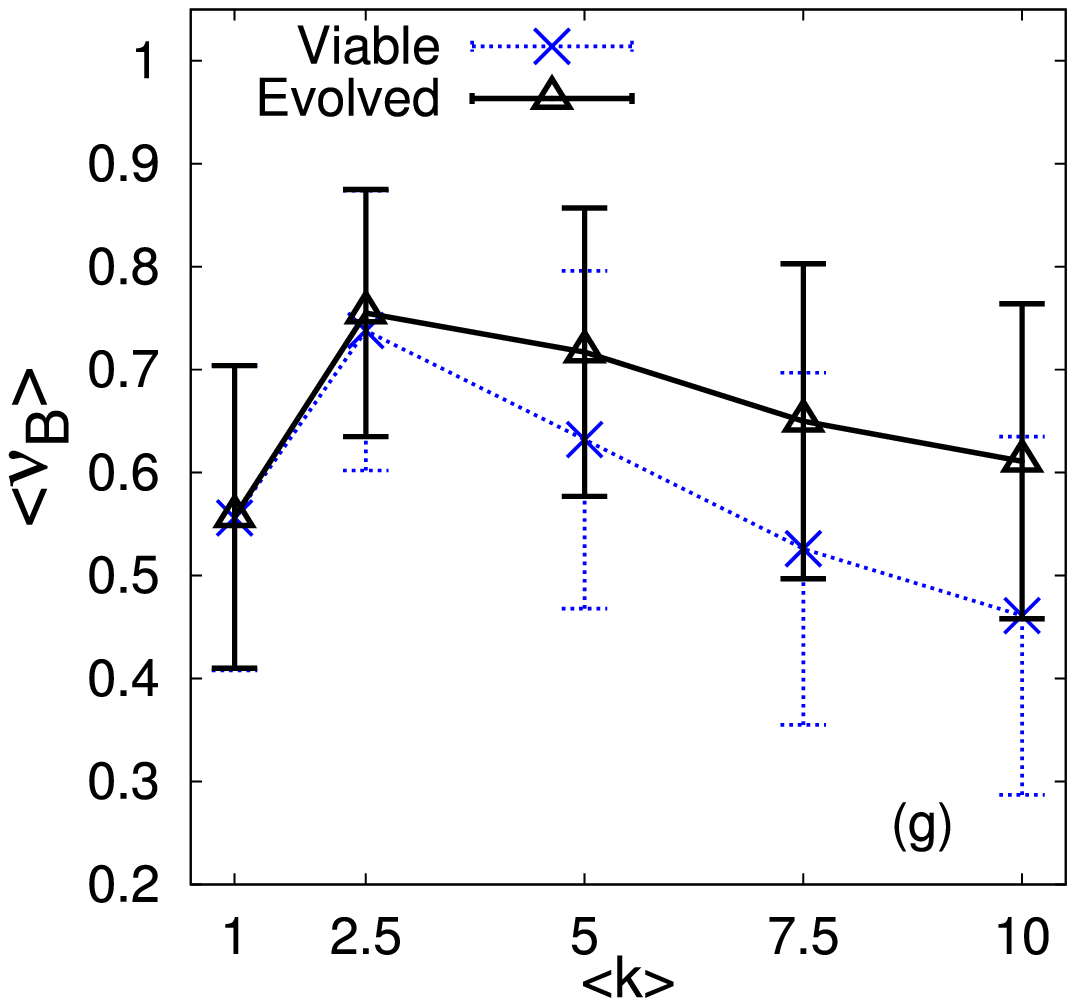}\hspace{.2truecm}\includegraphics[width=0.23\columnwidth]{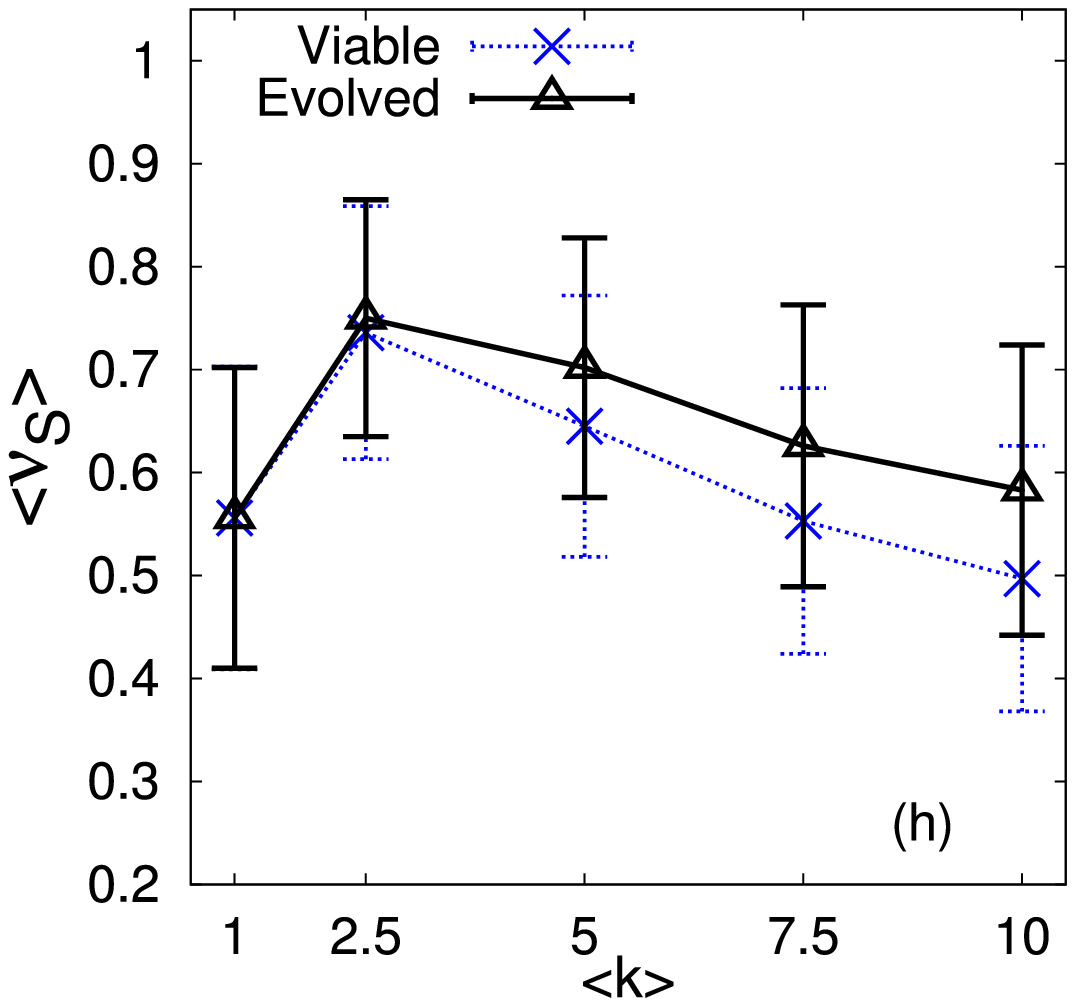}

\caption{\label{fig:before-and-after-plots} Mean mutational robustness ($R_{{\rm \mu}}),$
transient time $(\tau)$, normalized size of the principal basin $(B)$,
robustnesses of the trajectory $(\nu_{{\rm T}})$, initial state $(\nu_{{\rm s(0)}})$,
final state $(\nu_{{\rm s*}})$, principal basin $(\nu_{{\rm B}})$,
and state space $(\nu_{{\rm S}})$ before (viable) and after evolution (evolved)
 for $N=10$ and $\meank=1$, 2.5,
5, 7.5 and 10 averaged over 10,000 networks. The error bars are equal
to one standard deviation (not standard error). $B$ is normalized
by $\Omega/2=2^{N-1}$. The lines connecting the symbols are guides
to the eye.}
\end{figure}

As seen in Fig.~\ref{fig:before-and-after-plots}(a), networks of
all connectivities have similar values of mutational robustness,
$\Rmu$, before selection. After selection, however, the networks with
higher connectivity reach a much greater mean mutational robustness
\citep{Wagner:1996}. Sparsely connected networks do not show much
improvement because the low connectivity leaves little room for optimization.
Another significant change is seen in the transient time, $\tau$
(Figs.~\ref{fig:stats2}(g), (h), and (i) and \ref{fig:before-and-after-plots}(b))
\citep{Wagner:1996}. For viable networks, the transient time increases
with the connectivity. This is not surprising
since state spaces of densely connected networks typically have basins
with longer branches (Fig. \ref{fig:State-space-layout}(b)). However,
selection brings the mean transient time down to around 2, independent
of the connectivity of the network. Low transient time is one of the
properties of highly robust networks \citep{Ciliberti:2007}.

Mutations are not the only kind of perturbations that genetic regulatory
networks experience. All genetic systems are also exposed to noise
created by both internal an external sources. We measure robustness
to noise -- dynamical robustness -- using two parameters. The first
one is the dynamical robustness, $\nu$, of a gene expression state,
$\mathbf{s}$, to small perturbations (random one-bit flips), which
is simply the fraction of nearest neighbors in Hamming distance of $\mathbf{s}$ that
lie in the same basin as $\mathbf{s}$ \citep{Ciliberti:2007}. In
other words, $\nu$ is the probability that a random flip of a gene
on $\mathbf{s}$ yields a state that drains into the same attractor
as $\mathbf{s}$. We also employ the mean robustness of a set of states,
for instance, robustness of the gene expression trajectory, $\nu_{\mathrm{T}}$,
which is the average of $\nu$ over all states in the trajectory (including
$\initst$ and $\finst$). Similarly, the robustness of the principal
basin, $\nu_{\mathrm{B}}$, is the average of $\nu$ over all states
in the basin, and the robustness of the entire state space, $\nu_{\mathrm{S}}$,
is the average of $\nu$ over all possible states of the network.
(We usually omit the term {}``dynamical'' when we talk about robustness
of a state or a set of states since the mutational robustness of a
\emph{state} is not defined.)

The second parameter we use to estimate dynamical robustness is the
normalized size of the principal basin, $B$, which is the basin containing
both $\initst$ and $\finst$. The size of a basin is the total number
of states it contains. Therefore, $B$ is the fraction of all possible
perturbations to the trajectory that leave $\finst$ unchanged, i.e.,
the probability that flipping an arbitrary number of genes at random
in a state on the trajectory does not change the attractor that the
network settles into.

Clearly, the parameters we employ to measure robustness to noise are
quite different than conventional measures of stability, such as the
damage-spreading rate. This is because we are
interested in the endpoints of the gene expression trajectory, but
not in the exact path taken. Therefore, the relationship between the
damage spreading and robustness to noise is not trivial. In fact,
it is quite counterintuitive as explained below. 

As shown in Figs.~\ref{fig:stats1}(g), (h) and (i), densely connected
networks have a greater number of larger basins as their basin-size
distributions essentially follow a power law for about two decades.
The effect of the connectivity on $B$ is similar, as shown in Fig.~\ref{fig:before-and-after-plots}(c).
Networks with low connectivity have small principal basins both before
and after selection. More densely connected networks have larger $B$
before selection. We see a small increase in networks with $\meank=5$,
while fully connected networks ($\meank=10)$ increase their $\langle B\rangle$
significantly after evolution. However, compared to the change in
mutational robustness, the effect of selection on the size of the
principal basin is modest for highly connected networks.

Although greater connectivity increases the damage spreading
(sensitivity to small perturbations), the dynamical robustness of
the trajectory, $\nu_{{\rm T}}$, as well, increases with the connectivity
(Fig.~\ref{fig:before-and-after-plots}(d)). Networks with $\meank=1$
have a small $\langle\nu_{{\rm T}}\rangle$ both before and after selection.
The $\nu_{{\rm T}}$ curve for the viables has a maximum around $\meank=2.5$,
which also shows no improvement after selection. This indicates that
the networks with $\meank\approx2.5$ are intrinsically robust to
noise, but not evolvable. Viable networks with higher connectivity
have monotonically decreasing values of $\langle\nu_{{\rm T}}\rangle$,
but the selection improves their average robustness significantly,
above or up to the same level as the networks with $\meank=2.5$.
The robustnesses of the initial state, $\nu_{{\rm s(0)}}$, and the
final state, $\nu_{{\rm s*}},$ (Figs.~\ref{fig:before-and-after-plots}(e)
and (f)) follow the same trend. The robustnesses of the principal
basin, $\nu_{{\rm B}}$, and state space, $\nu_{{\rm S}}$, (Figs.~\ref{fig:before-and-after-plots}(g)
and (h)) do not show as much improvement with selection. This implies
that the effect of selection is more local (on the specific gene-expression
trajectory) than global in terms of the change in dynamics. Our results
above agree with the findings by \citet{Ciliberti:2007}, who used very similar
parameters to measure robustness to noise. 

We note that most of the robustness measures discussed above do not
increase monotonically with $\langle k\rangle$ for the viable networks
(i.e., before selection). For example, $\langle R_{{\rm \mu}}\rangle$
for viable networks has a maximum around $\langle k\rangle=5$ (Fig.~\ref{fig:before-and-after-plots}(a)).
This probably indicates that these quantities depend on more than
one parameter (like the principal basin size and the magnitude of the damage spreading), some increasing and some decreasing with increasing $\langle k\rangle$.
For the evolved networks, however, $\langle R_{{\rm \mu}}\rangle$,
$\langle B\rangle$, and $\langle\nu_{{\rm T}}\rangle$ increase monotonically
with $\langle k\rangle,$ at least for the range tested in this paper.
This indicates that networks of higher connectivity are more evolvable
when mutational robustness is considered. However, these networks
are more chaotic on average compared to the ones of low connectivity,
regardless of selection. 

\begin{figure}
\includegraphics[width=0.23\columnwidth]{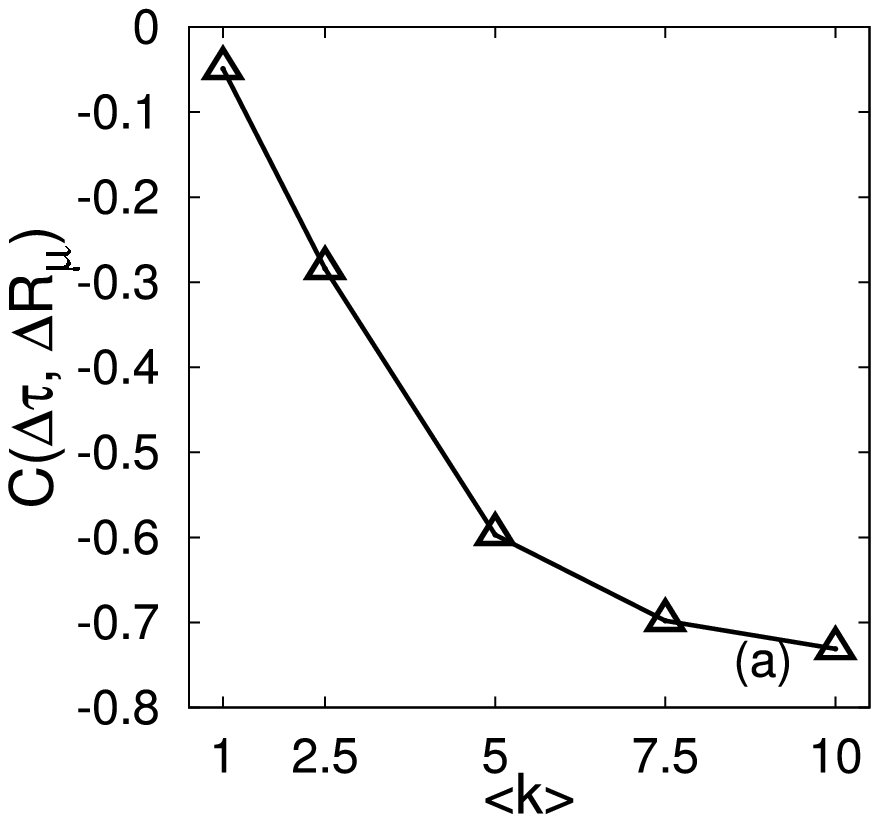}\hspace{.2truecm}\includegraphics[width=0.23\columnwidth]{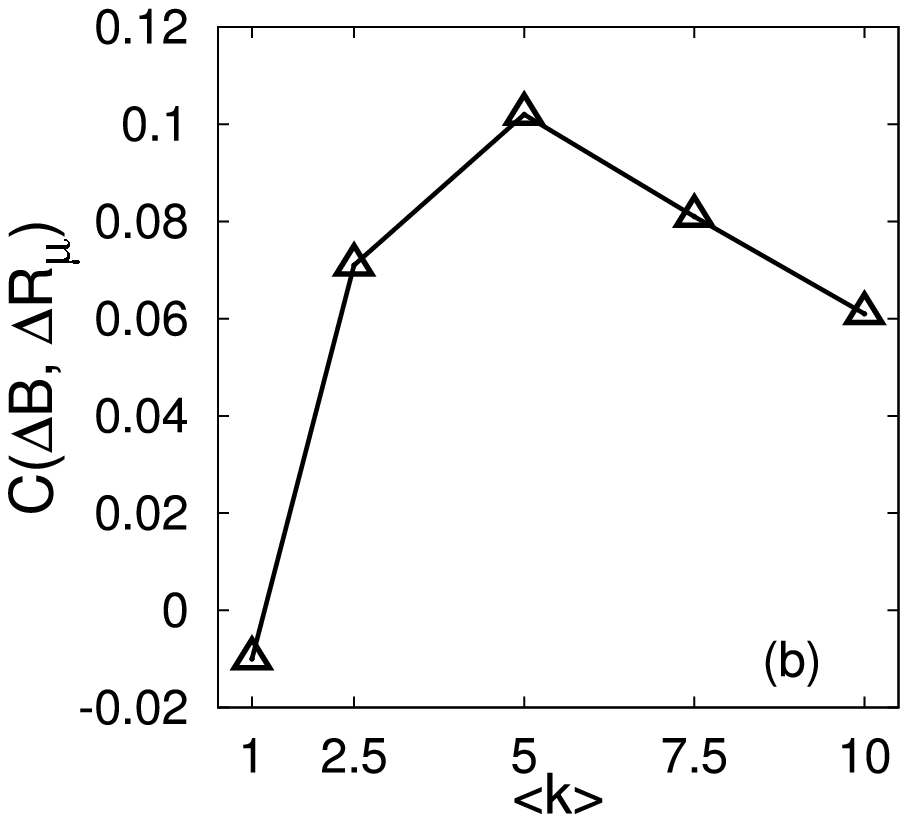}\hspace{.2truecm}\includegraphics[width=0.23\columnwidth]{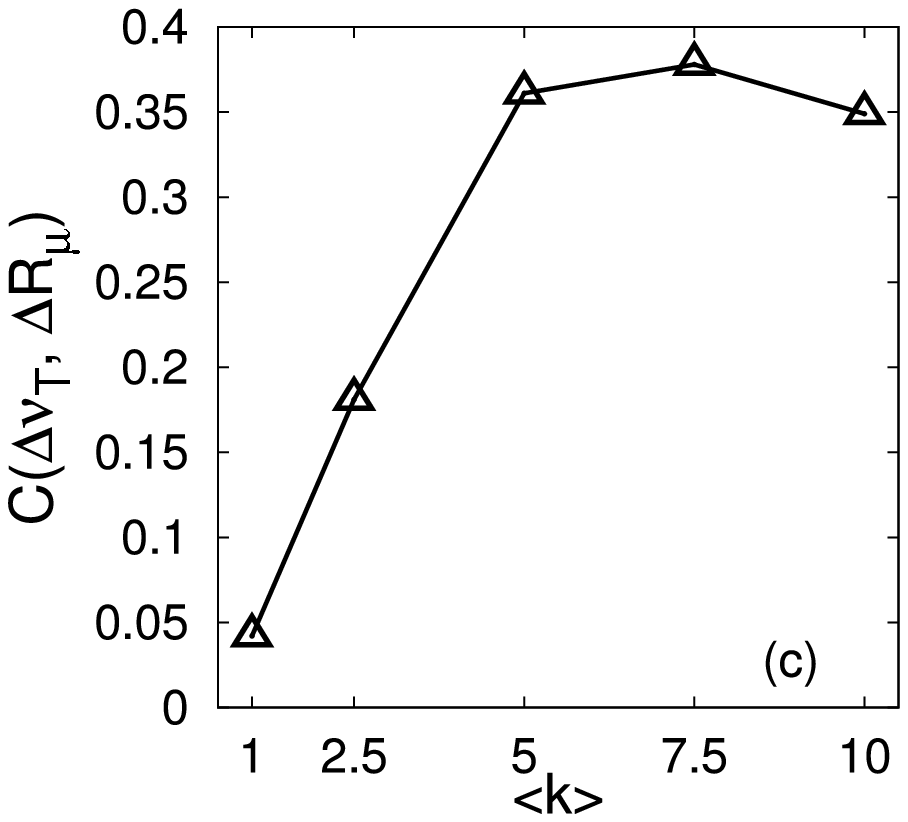}\hspace{.2truecm}\includegraphics[width=0.23\columnwidth]{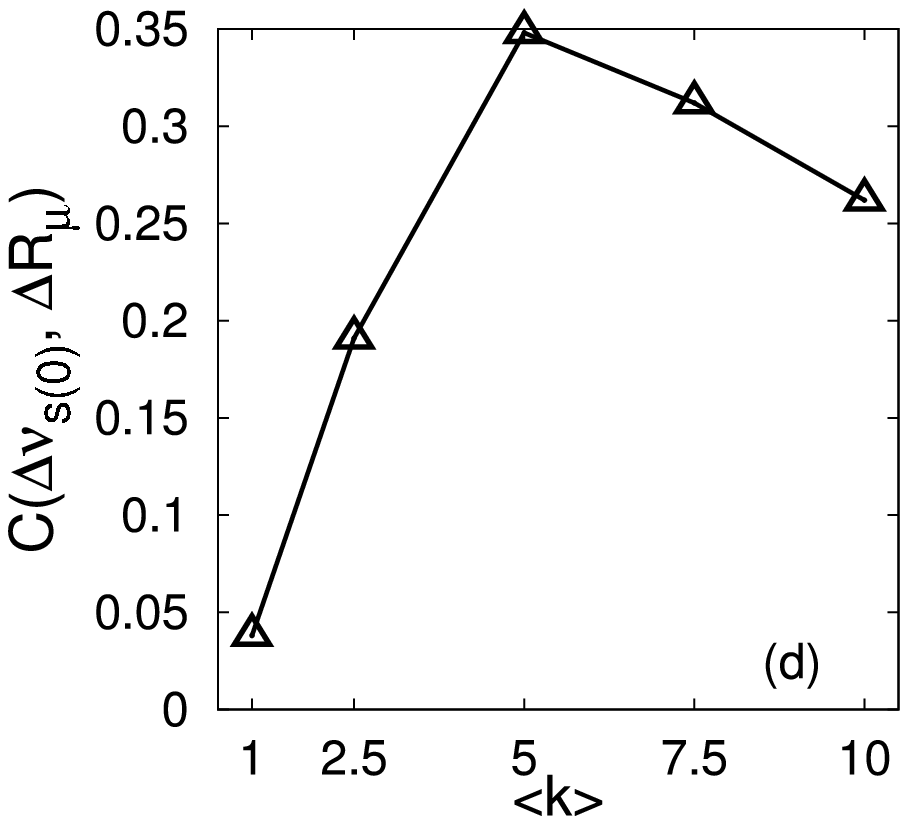}

\includegraphics[width=0.23\columnwidth]{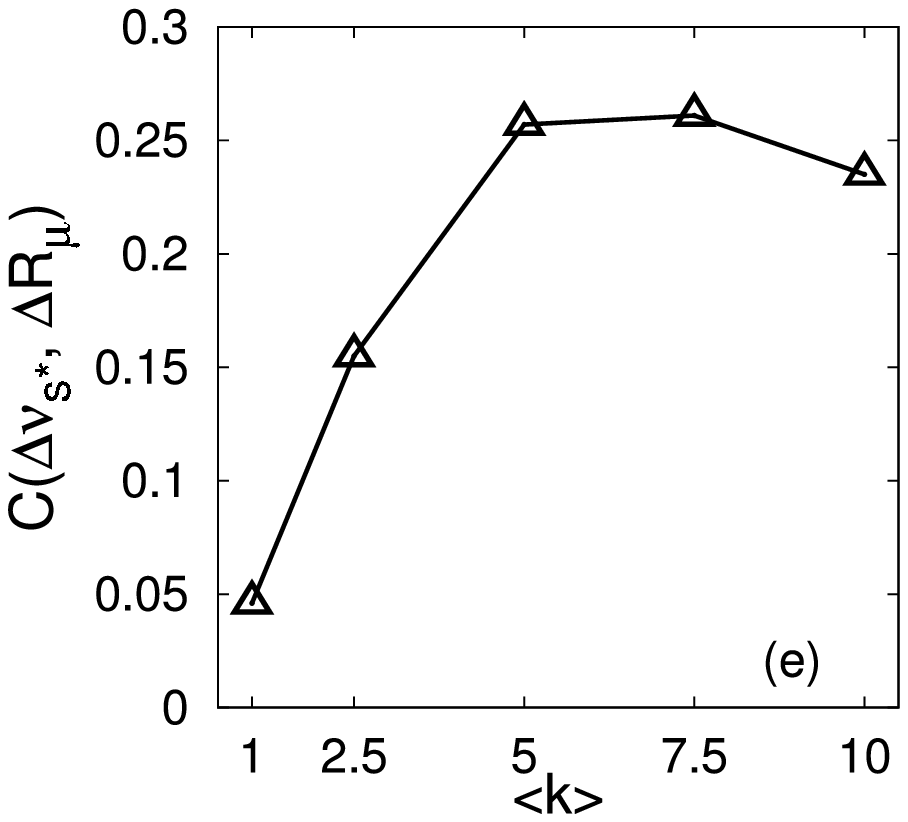}\hspace{.2truecm}\includegraphics[width=0.23\columnwidth]{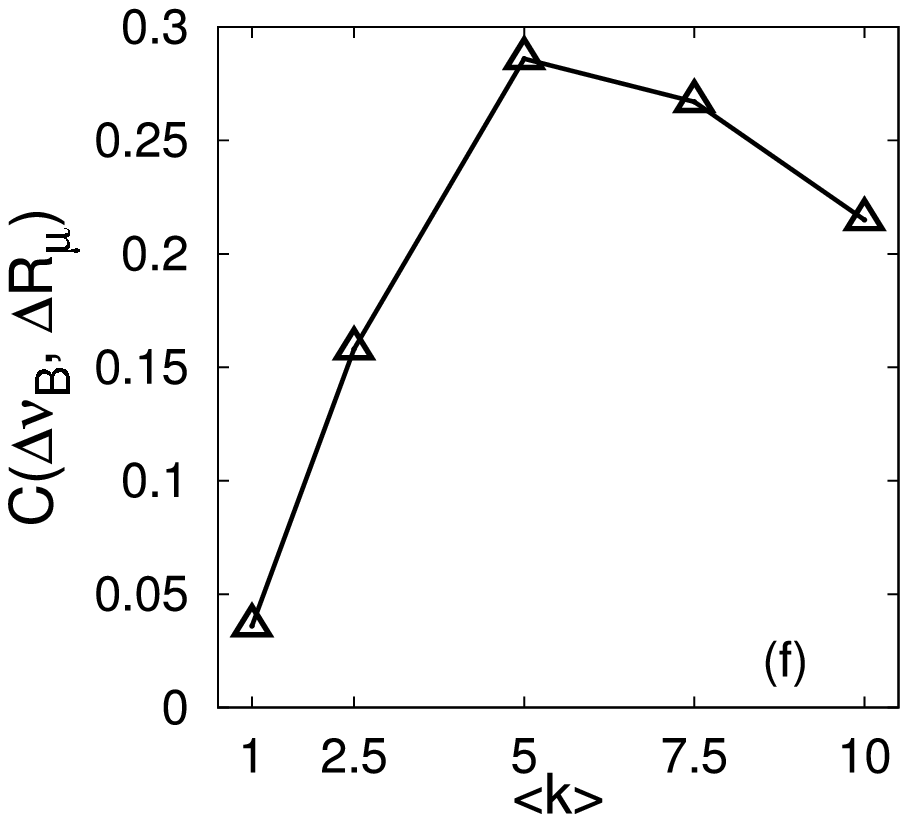}\hspace{.2truecm}\includegraphics[width=0.23\columnwidth]{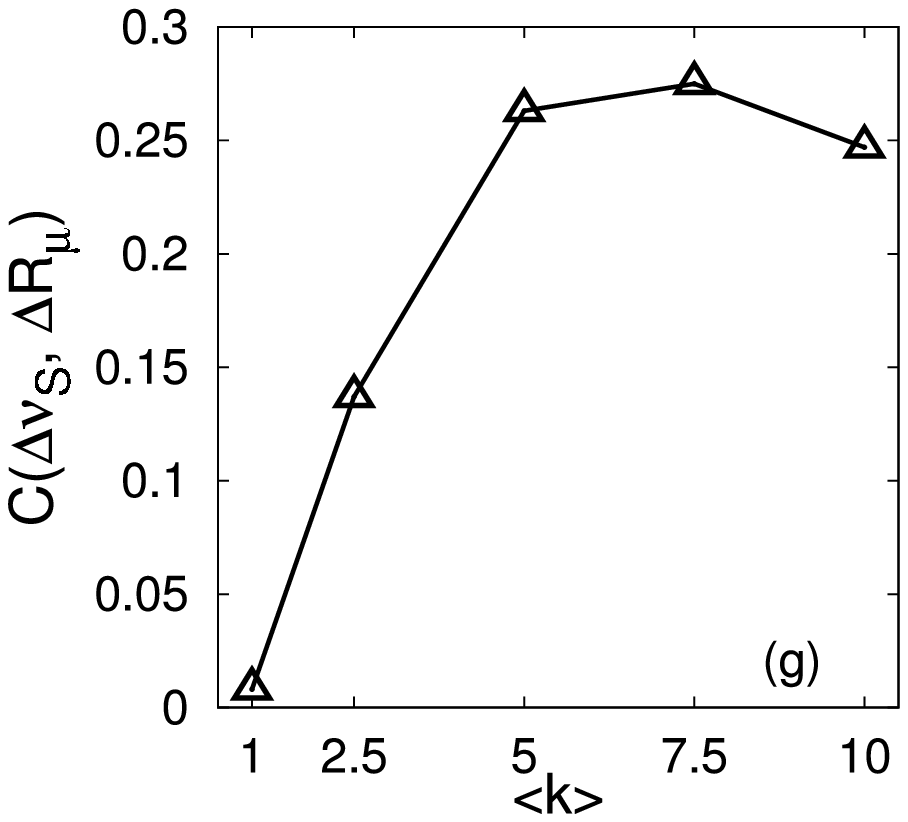}\hspace{.2truecm}\includegraphics[width=0.23\columnwidth]{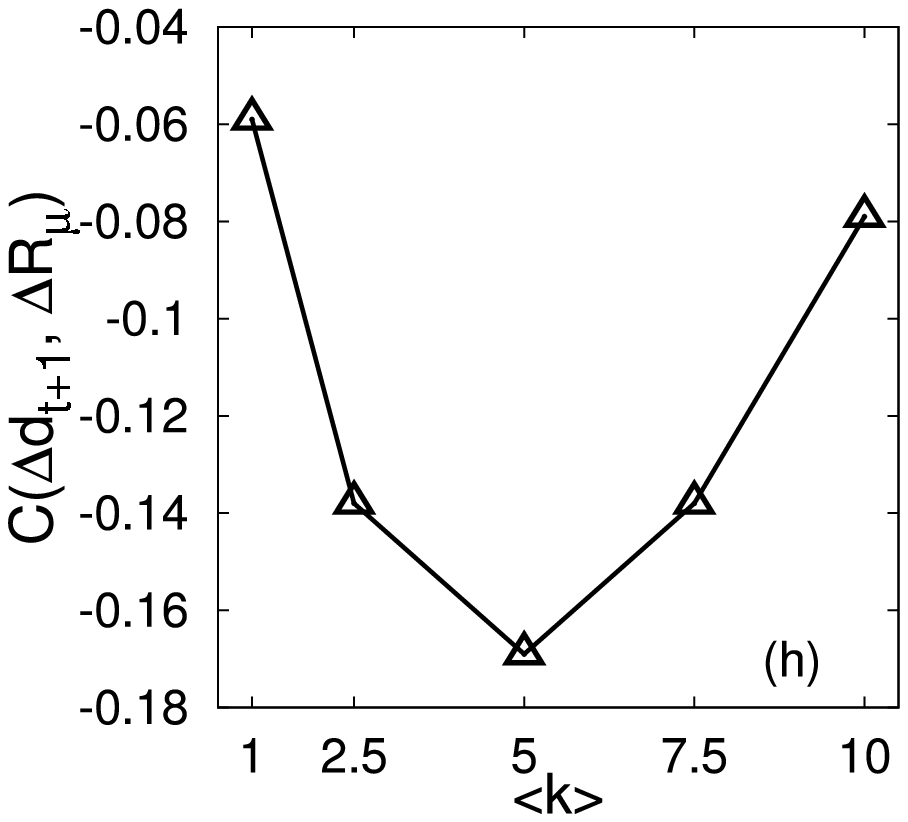}

\caption{Correlations between change in state-space parameters and change
in robustness, calculated using 10,000 individual evolved networks
with $N=10$ and $\meank=1$, 2.5, 5, 7.5, and 10. The $P$-value is much smaller than 0.01 for
all correlation coefficients with an absolute value above 0.03. $\Delta\tau$,
$\Delta B$, $\Delta\nu_{{\rm T}}$, $\Delta\nu_{{\rm s(0)}}$, $\Delta\nu_{{\rm s*}}$,
$\Delta\nu_{{\rm B}}$, $\Delta\nu_{{\rm S}},$ and $\Delta d_{t+1}$
denote changes in normalized size of the transient time, size of the
principal basin, robustnesses of the trajectory, initial state, final
state, principal basin, and state space, and damage spreading, 
respectively. The dynamical robustness of the trajectory includes
the contribution from the final state. The lines connecting the symbols
are guides to the eye. See text for details.\label{fig:Correlations}}
\end{figure}
%
{}

\subsection{Correlations between Change in Mutational Robustness and Other Parameters
\label{sub:What-is-Being-Optimized}}

The recent theoretical and computational work on gene regulatory networks
\citep{Ciliberti:2007,Kaneko:2007PLOSONE} indicates that there is
a strong link between mutational robustness and robustness to noise.
In order to see the relation between these two quantities, we
calculated the correlation coefficients between
the change in mutational robustness ($\Delta R_{{\rm \mu}}$) and
changes in different measures of robustness to noise, as well as transient
time and the damage spreading as functions of linkage density. 

Changes in the transient time and mutational robustness are highly
correlated in agreement with the earlier studies \citep{Wagner:1996}.
The principal basin size does not seem to have much effect on $R_{{\rm \mu}}$,
as the correlation between their changes is quite weak (Fig. \ref{fig:Correlations}(b)).
Although a larger principal basin size means greater robustness to
noise, this result suggests that it is not strongly selected for.
Changes in robustnesses of the initial state, final state, principal
basin, and state space are also correlated with $\Delta R_{{\rm \mu}}$
(Figs.~\ref{fig:Correlations}(c), (d), (e), and (f)). The weak correlation
between the changes in the damage spreading and mutational robustness
(Fig.~\ref{fig:Correlations}(h)) imply that it is not the change
in dynamical behavior that brings mutational robustness.

\section{Discussion\label{sec:Discussion}}

In this paper, we have analyzed changes in state-space properties of model
genetic regulatory networks under selection for an optimal phenotype.
Both numerical stability analysis and the state-space statistics indicate
that the difference between the global dynamical properties of mutationally
robust networks that have undergone selection and their random ancestors
are quite small. Furthermore, the correlation between the changes
in the damage spreading and the mutational robustness is weak.
Therefore, changes in the global dynamical properties do not seem
to be responsible for the increase in mutational robustness after
selection. 

Dynamics of many random threshold networks, as well as Random Boolean
Networks, depend largely on their connectivity distributions. 
Variants of the Random Threshold Networks used in this paper have
been shown to have a chaotic phase above $\langle k\rangle\approx2$, depending
on the model details \citep{Kurten:1988a,Rohlf:2002}. Essentially, RTNs have a chaotic phase for sufficiently large $\meank$  \citep{Rohlf:2002,Kurten:1988,Kurten:1988a}
just as RBNs \citep{KAUF93,AldanaReview,DrosselRBNReview}. 

The connectivity of a network affects the evolvability of its mutational
robustness, as well as its dynamical character. For viable (essentially
random) networks, the mutational robustness is very similar for all
connectivities \citep{Wagner:1996}. For the evolved networks, however,
it increases monotonically with increasing connectivity, creating
drastic differences for large $\meank$. The dynamical robustness
of the gene-expression trajectory, the initial state, and the final
state follow similar trends. These results clearly indicate that the stability
of the system as measured by the damage spreading does not capture
its dynamical characteristics in this context. 

As pointed out in a recent paper by  \citet{Ciliberti:2007},
selection decreases the transient time by picking the {}``proper''
interaction constants to construct a shorter (or direct) path from
$\initst$ to $\finst$. Both \citet{Ciliberti:2007}
and \citet{Kaneko:2007PLOSONE} stated that
mutationally robust networks are the ones that have found a path for the
gene expression trajectory at a safe distance from the basin boundary,
so that small perturbations cannot kick them into a different basin.
Thus, even if the selection operates only on the stability of the stationary
gene-expression pattern, robustness to mutations intrinsically requires
stability of the gene-expression trajectory against small perturbations.
There is also some experimental evidence supporting the association
between genetic and non-genetic change \citep{WagnerBook,Stearns:1995}.
Although it might seem like robustness to noise evolves as a by-product
of robustness to mutations, the converse case is also true \citep{Kaneko:2007PLOSONE}.
Thus, robustness to noise and robustness to mutations seem to evolve
mutually when certain conditions are met \citep{Kaneko:2007PLOSONE}.

The analysis of the cell-cycle regulatory network of budding yeast
provides further evidence for the chaotic behavior of gene networks
\citep{Li:2006,Ganguli:2007}. The simplified form of this network
has 11 nodes (genes or proteins), one checkpoint (an external input,
in this case, the cell size), and 34 links including self-degrading
interactions. Using a dynamical model similar to the one used in this
paper, \citet{Li:2006} \emph{}showed  that the stationary
state of the network has a basin occupying 86\% of the state space.
\citet{Ganguli:2007} studied an ensemble of networks
that can perform the same function, i.e., the 12-step sequence of
transitions in the expression trajectory, and found that these functional
networks have larger basins for the stationary state (consequently,
broadly distributed basin sizes), fewer attractors, longer transient
times, and a larger damage-spreading rate compared to their randomized
counterparts. They concluded that those dynamical features emerge
due to the functional constraints on the network. Here, we showed
that those features, which are signs of chaotic dynamics, can arise under the
presence of structural perturbations (mutations) if the connectivity
of the network is large enough, even when the constraints on the
function are minimal, i.e., when the only selection is on the phenotype.
Although the length of the gene-expression trajectory of the yeast
cell-cycle network needs further explanation in terms of mutational
robustness, it appears like chaotic dynamics may be a design principle
underlying seemingly {}``boring'' and ordered behavior generally
seen in models of gene regulatory networks, where a simple cascade
of expression terminates at a stationary state \citep{DrosselSzejka}. %
{}

The effect of network topology on evolvability of robustness is another
aspect of the problem, which we do not discuss in this paper \citep{Jeong:2000,Jeong:2001,Oikonomou:2006}.
However, we would like to point out that recent studies indicate that
certain topological features, such as connectivity, are not very crucial
in determining the response of cellular networks to genetic or non-genetic
change \citep{Hahn:2004,Siegal:2007}, and there may be other factors
shaping their topological structure \citep{Balcan:2007PLOSONE,Socolar:2006}. 

Our results also imply that the {}``life at the edge of chaos''
hypothesis \citep{KAUF93}, which suggests adaptability (evolvability)
is maximized in the critical regime does not seem to be necessary,
at least not to explain evolvability of robustness to mutations and noise.
Indeed, recent studies concerning dynamics of genetic regulatory networks
do not indicate any special feature brought by criticality \citep{Aldana:2003}.
The {}``edge of chaos'' concept was primarily developed to describe
the phase in which cellular automata can perform universal computation
\citep{Langton:1990,Mitchell:1993}, and it may not be related to dynamics
and evolution of biochemical pathways as was once thought.

To summarize, our study indicates that conventional measures of stability
may not be very informative about robustness to mutations or noise
in gene regulatory networks when one considers the steady gene-expression
pattern as the robust feature of the network. Also, the dynamics underlying 
the simple gene-expression trajectories can be very rich, 
reflecting a complex state-space structure.

\section*{Supplementary material}
The online version of this article contains supplementary material: Linearly binned 
histograms of the data represented in Figs.~\ref{fig:stats1} and \ref{fig:stats2}, 
probability densities for the Hamming distance between the initial and final states, 
the magnitude of the damage spreading before and after evolution, and probability densities for the latter.

\section*{Acknowledgments}
We thank T. F. Hansen, S. Bornholdt, G. Brown, D. Balcan, B. Uzuno\u glu, P.
Oikonomou, and A. Pagnani for helpful discussions. This research was
supported by U.S. National Science Foundation Grant Nos. DMR-0240078
and DMR-0444051, and by Florida State University through the School
of Computational Science, the Center for Materials Research and Technology,
and the National High Magnetic Field Laboratory.

\end{document}